\documentclass[a4paper,11pt]{article}
\pdfoutput=1
\usepackage{jheppub}

\usepackage{graphicx}
\usepackage{amssymb}
\usepackage{amsthm}
\usepackage{amsmath}
\usepackage{color}
\usepackage{xspace}
\usepackage{preprintcover}
\usepackage{atlasphysics}
\usepackage{datetime}
\usepackage{nicefrac}
\usepackage{rotating}
\usepackage{multirow}
\usepackage{relsize}
\usepackage[section]{placeins}

\newcommand{\powheg} {{\sc Powheg-Box}\xspace}
\newcommand{\pythia} {{\sc Pythia6}\xspace}
\newcommand{\pythiapp} {{\sc Pythia8}\xspace}
\newcommand{\prospino} {{\sc Prospino2}\xspace}

\newcommand{\herwigpp} {{\sc Herwig++}\xspace}
\newcommand{\madgraph} {{\sc MadGraph5}\xspace}
\newcommand{\bridge} {{\sc Bridge}\xspace}

\newcommand{\geant} {{\sc Geant4}\xspace}

\newcommand{\rhad} {\mbox{$R$-hadron}\xspace}
\newcommand{\rhads} {\mbox{$R$-hadrons}\xspace}
\newcommand{\lsusy} {\mbox{LeptoSUSY}\xspace}
\newcommand{\lumi} {19.1~\ifb\xspace}
\newcommand{\lumiZmumu} {19.8~\ifb\xspace}
\renewcommand{\pt} {\ensuremath{p_{\mathrm{T}}}\xspace}

\newcommand{\zmumu} {\ensuremath{Z \to \mu\mu}\xspace}
\newcommand{\etmiss} {\ensuremath{E_{\mathrm{T}}^\mathrm{miss}}\xspace}
\newcommand{\ptmiss} {\ensuremath{\vec{p}_{\mathrm{T}}^\mathrm{~miss}}\xspace}
\newcommand{\CLs} {\ensuremath{CL_{s}}\xspace}

\renewcommand{\MeV} {\ensuremath{\mathrm{{Me\kern -0.1em V}}}\xspace}
\renewcommand{\GeV} {\ensuremath{\mathrm{{Ge\kern -0.1em V}}}\xspace}
\renewcommand{\TeV} {\ensuremath{\mathrm{{Te\kern -0.1em V}}}\xspace}
\renewcommand{\stau} {\ensuremath{\tilde{\tau}}\xspace}
\newcommand{\MeVgcm} {\ensuremath{\MeV\mathrm{g}^{-1}\mathrm{cm}^{2}}\xspace}
\newcommand{\dedx} {{\ensuremath{\mathrm{d}E/\mathrm{d}x}}\xspace}

\newcommand{\srslcc} {{\tt SR-SL-2C}\xspace}
\newcommand{\srslc} {{\tt SR-SL-1C}\xspace}

\newcommand{\srchcc} {{\tt SR-CH-2C}\xspace}
\newcommand{\srchcl} {{\tt SR-CH-1LC}\xspace}
\newcommand{\srchc} {{\tt SR-CH-1C}\xspace}
\newcommand{\srrhma} {{\tt SR-RH-MA}\xspace}
\newcommand{\srrhfd} {{\tt SR-RH-FD}\xspace}
\newcommand{\combined} {{\smaller \textsc{combined}}\xspace}
\newcommand{\idcalo} {{\smaller \textsc{id+calorimeter}}\xspace}

\newcommand{\hide}[1] {}

\newcommand{\limitleptosquark}{1500\xspace}
\newcommand{\limitleptogluino}{1360\xspace}
\newcommand{\limitcharginos}{620\xspace}
\newcommand{\limitgluinofull}{1270\xspace}
\newcommand{\limitsbottomfull}{845\xspace}
\newcommand{\limitstopfull}{900\xspace}
\newcommand{\limitgluinoagno}{1260\xspace}
\newcommand{\limitsbottomagno}{835\xspace}
\newcommand{\limitstopagno}{870\xspace}
\newcommand{\gmsbAllExclusion}{440, 440, 430, 410, 385\xspace}

\newcommand{\puttitle}{Searches for heavy long-lived charged particles with the ATLAS detector in proton--proton collisions at $\sqrt{s} = 8~\TeV$}
\newcommand{\putabstract}{Searches for heavy long-lived charged particles are performed using a data sample of \lumi from proton--proton collisions at a centre-of-mass energy of $\sqrt{s} = 8~\TeV$ collected by the ATLAS detector at the Large Hadron Collider. No excess is observed above the estimated background and limits are placed on the mass of long-lived particles in various supersymmetric models. 
Long-lived tau sleptons in models with gauge-mediated symmetry breaking are excluded up to masses between 440 and 385~\GeV for $\tan\beta$ between 10 and 50, with a 290~\GeV limit in the case where only direct tau slepton production is considered. In the context of simplified \lsusy models, where sleptons are stable and have a mass of 300~\GeV, squark and gluino masses are excluded up to a mass of \limitleptosquark and \limitleptogluino~\GeV, respectively. Directly produced charginos, in simplified models where they are nearly degenerate to the lightest neutralino, are excluded up to a mass of \limitcharginos~\GeV. \rhads, composites containing a gluino, bottom squark or top squark, are excluded up to a mass of \limitgluinofull, \limitsbottomfull and \limitstopfull~\GeV, respectively, using the full detector; and up to a mass of \limitgluinoagno, \limitsbottomagno and \limitstopagno~\GeV using an approach disregarding information from the muon spectrometer.}

\newdateformat{vdate}{\twodigit{\THEYEAR}-\twodigit{\THEMONTH}-\twodigit{\THEDAY}}
\newtimeformat{vtime}{\twodigit{\THEHOUR}-\twodigit{\THEMINUTE}}

\title{\boldmath \puttitle}
\collaboration{ATLAS Collaboration}

\abstract{\putabstract}

\arxivnumber{1411.6795}

\PreprintCoverPaperTitle{\puttitle}

\PreprintIdNumber{CERN-PH-EP-2014-252}

\PreprintCoverAbstract{\putabstract}

\PreprintJournalName{JHEP}

\begin{document}
\maketitle
\flushbottom
\setlength{\parindent}{0cm}
\setlength{\parskip}{0.25cm}

\section{Introduction} \label{sec:introduction}

Heavy long-lived particles (LLP) are predicted in a range of extensions of the Standard Model (SM)~\cite{Fairbairn:2006gg}. $R$-parity-conserving supersymmetry (SUSY)~\cite{Miyazawa:1966, Ramond:1971gb, Golfand:1971iw, Neveu:1971rx, Neveu:1971iv, Gervais:1971ji, Volkov:1973ix, Wess:1973kz, Wess:1974tw, Fayet:1976et, Fayet:1977yc, Farrar:1978xj, Fayet:1979sa, Dimopoulos:1981zb} models, such as split SUSY~\cite{ArkaniHamed:2004fb, ArkaniHamed:2004yi}, gauge-mediated SUSY breaking (GMSB)~\cite{Dine:1981gu, AlvarezGaume:1981wy, Nappi:1982hm, Dine:1993yw, Dine:1994vc, Dine:1995ag, Kolda:1997wt, Raby:1997pb} and \lsusy~\cite{DeSimone:2009ws, DeSimone:lgm}, as well as other scenarios such as universal extra dimensions~\cite{Appelquist:2000nn} and leptoquark extensions~\cite{Friberg:1997nn}, allow for a variety of LLP states stable enough to be directly identified by the ATLAS detector. These states include long-lived super-partners of the leptons, quarks and gluons; sleptons ($\tilde{\ell}$), squarks ($\tilde{q}$) and gluinos ($\tilde{g}$), respectively; as well as charginos ($\tilde{\chi}^{\pm}_{1,2}$), which together with neutralinos ($\tilde{\chi}^{0}_{1-4}$) are a mixture of super-partners of the Higgs and $W$/$Z$ bosons, known as Higgsinos, winos and binos.

When travelling with a speed measurably slower than the speed of light, charged particles can be identified and their mass ($m$) determined from their measured speed ($\beta$) and momentum ($p$), using the relation $m=p/\beta\gamma$, where $\gamma$ is the relativistic Lorentz factor. Three different searches are presented in this article, using time-of-flight (TOF) to measure $\beta$ and specific ionisation energy loss (\dedx), to measure $\beta\gamma$.

The searches are based almost entirely on the characteristics of the LLP itself, but are further optimised for the different experimental signatures of sleptons, charginos and composite colourless states of a squark or gluino together with light SM quarks or gluons, called \rhads.

Long-lived charged sleptons would interact like muons, releasing energy by ionisation as they pass through the ATLAS detector. A search for long-lived sleptons identified in both the inner detector (ID) and in the muon spectrometer (MS) is therefore performed (``slepton search''). The search is optimised for GMSB and \lsusy models. In the former, the gravitino is the lightest supersymmetric particle (LSP) and the light tau slepton ($\stau_1$) is the long-lived, next-to-lightest supersymmetric particle (NLSP). The $\stau_1$, the lightest \stau mass eigenstate resulting from the mixture of right-handed and left-handed super-partners of the $\tau$ lepton, is predominantly the partner of the right-handed lepton in all models considered here. In addition to GMSB production, results are also interpreted for the case of direct pair production of charged sleptons, independently of the mass spectrum of other SUSY particles. The recent discovery of the Higgs boson with a mass of about 125~\GeV~\cite{Aad:2012tfa, Chatrchyan:2012ufa} disfavours minimal GMSB within reach of the Large Hadron Collider (LHC). For the Higgs boson to have such mass, the top squark mass would have to be several \TeV, and in GMSB the slepton masses are strictly related to the squark masses. However, modifications to minimal GMSB can easily accommodate the observed Higgs mass without changing the sparticle masses~\cite{Abdullah, EvansIbeYanigida, EvansIbeShiraiYanigida}. The \lsusy models, characterised by final states with high multiplicity of leptons and jets, are studied in the context of a simplified model, where all the neutralinos and charginos are decoupled with the exception of the $\tilde\chi_1^0$, and the sleptons are long-lived and degenerate, with a mass set to 300~\GeV, a value motivated by exclusion limits of previous searches~\cite{Aad:2012pra}. In these models a substantial fraction of the events would contain two LLP candidates, a feature also used to discriminate signal from background.

Charginos can be long-lived in scenarios where the LSP is a nearly pure neutral wino and is mass-degenerate with the charged wino. The chargino signature in the detector would be the same as for a slepton, but the dominant production is in chargino--neutralino ($\tilde{\chi}^{\pm}_1\tilde{\chi}^{0}_1$) pairs, where the neutralino leaves the apparatus undetected. As a result, the event would have one LLP and significant missing transverse momentum (\ptmiss, with magnitude denoted by \etmiss). This signature is pursued in a dedicated ``chargino search''.

Coloured LLPs ($\tilde{q}$ and $\tilde{g}$) would hadronise forming \rhads, bound states composed of the LLP and light SM quarks or gluons. They may emerge as charged or neutral states from the $pp$ collision and be converted to a state with a different charge by interactions with the detector material, and thus arrive as neutral, charged or doubly charged particles in the MS. Searches for \rhads are performed following two different approaches: using all available detector information (``full-detector \rhad search''), or disregarding all information from the MS (``MS-agnostic \rhad search''). The latter case is independent of the modelling of \rhad interactions with material in the calorimeters.

Previous collider searches for charged LLPs have been performed at LEP~\cite{Barate:1997dr,Abreu:2000tn,Achard:2001qw,Abbiendi:2003yd}, HERA~\cite{Aktas:2004pq}, the Tevatron~\cite{Abazov:2007ht,Aaltonen:2009kea,Abazov:2011pf}, and the LHC~\cite{Aad:2012pra,Chatrchyan:2013oca,Aad:2013gva}.

\section{Data and simulated samples}

The work presented in this article is based on \lumi of $pp$ collision data collected at a centre-of-mass energy $\sqrt{s} = 8~\TeV$ in 2012. Events are selected online by trigger requirements either on the presence of muons or large \etmiss. Events collected during times when a problem was present in one of the relevant sub-detectors are later rejected offline. A separate stream of \lumiZmumu $pp$ collision data and Monte Carlo (MC) simulation \zmumu samples are used for timing resolution studies. Simulated signal samples are used to study the expected signal behaviour and to set limits.

All MC simulation samples are passed through a detector simulation~\cite{Aad:1267853} based on \geant~\cite{Agostinelli:2002hh} and a model of the detector electronics. The effect of multiple $pp$ interactions in the same or a nearby bunch crossing (pile-up) is taken into account by overlaying additional minimum-bias collision events simulated using \pythiapp~\cite{Sjostrand:2007gs} v.\ 8.170 and reweighting the distribution of the average number of interactions per bunch crossing in MC simulation to that observed in data. All events are subsequently processed using the same reconstruction algorithms and analysis chain as the data. 

The GMSB samples are generated, using \herwigpp~\cite{Bahr:2008pv} v.\ 2.5.2 along with the UEEE3~\cite{Gieseke:2012ft} tune and the CTEQ6L1~\cite{Pumplin:2002vw} parton distribution function (PDF) set, with the following model parameters: number of super-multiplets in the messenger sector, $N_5=3$, messenger mass scale, $m_{\rm{messenger}}=250~\TeV$, sign of the Higgsino mass parameter, $\rm{sign}(\mu)=1$, and $C_\mathrm{grav}$, the scale factor for the gravitino mass which determines the NLSP lifetime, set to 5000 to ensure that the NLSP does not decay inside the detector. The ratio of the vacuum expectation values of the two Higgs doublets ($\tan\beta$) is varied between 10 and 50. The SUSY-breaking scale ($\Lambda$) is chosen between 80 and 160~\TeV and the corresponding $\stau_1$ masses vary from 175 to 510~\GeV, in order to cover the regions of parameter space accessible to this analysis and not excluded by previous searches. The masses of the right-handed $\tilde{e}$ (or $\tilde{\mu}$) are larger than that of $\stau_1$ by 2.7--93~\GeV for $\tan\beta$ values between 10 and 50. The corresponding lightest neutralino ($\tilde{\chi}^{0}_{1}$) mass varies from 328 to 709~\GeV as a function of $\Lambda$ and is independent of $\tan\beta$. The lightest chargino ($\tilde{\chi}^{\pm}_{1}$) mass varies from 540 to 940~\GeV, and is 210 to 260~\GeV higher than the neutralino mass, with a small dependence on $\tan\beta$. The dependence of the mass splitting between the chargino and lightest neutralino on $\tan\beta$ varies from 1\% at $\Lambda=80$~\TeV to 3\% at $\Lambda=160$~\TeV.

The \lsusy samples are simulated in \madgraph~\cite{Alwall:2011uj} v.\ 1.5.4 using the CTEQ6L1 PDF set, with \bridge~\cite{Meade:2007js} v.\ 2.24 used for decaying the squarks, and \pythiapp v.\ 8.170 along with the AU2~\cite{ATL-PHYS-PUB-2011-008,ATL-PHYS-PUB-2011-009} tune for parton showering. The sleptons are long-lived and set to be degenerate with a mass of 300~\GeV. The third-generation squarks are assumed to be very heavy (10~\TeV). 
The masses of the first- and second-generation squarks (gluinos) are varied between 600~\GeV and 3~\TeV (950~\GeV and 3~\TeV) assuming a fixed mass of the $\tilde{\chi}^{0}_{1}$ of 400~\GeV.

Samples of long-lived charginos are generated using \herwigpp v.\ 2.6.3 along with the UEEE3 tune and the CTEQ6L1 PDF set, according to simplified models where the lightest chargino and lightest neutralino are nearly degenerate, and the chargino is the LLP. Starting from a self-consistent model with a chargino/neutralino mass of about 658~\GeV (140~\MeV mass splitting), the simplified version is obtained by moving the chargino and neutralino masses up and down in a range between 100 and 800~\GeV, keeping the mass splitting constant. In addition, the chargino is forced to remain stable and the other particle masses are set to values too high to be produced at the LHC. Production of $\tilde{\chi}^{\pm}_{1}\tilde{\chi}^{\mp}_{1}$ ($\tilde{\chi}^{\pm}_{1}$$\tilde{\chi}^{0}_{1}$) constitutes about one third (two thirds) of the events generated in these samples.

For the \rhad samples, pair production of gluinos, bottom squarks (sbottoms) and top squarks (stops) is simulated in \pythia~\cite{Sjostrand:2006za} v.\ 6.4.27, incorporating specialised hadronisation routines~\cite{rasmusthesis,Kraan:2004tz} to produce final states containing \rhads~\cite{Aad:2011yf}, along with the AUET2B~\cite{ATL-PHYS-PUB-2011-014} tune and the CTEQ6L1 PDF set. Interactions of \rhads with matter are handled by dedicated routines for \geant based on different scattering models with alternative assumptions~\cite{Mackeprang:2009ad}. The model for gluino \rhad interactions, using a gluino-ball fraction of ten percent, is referred to as the generic model. For sbottom and stop \rhads a triple Regge interaction model is assumed.

Samples of \zmumu events are simulated using \powheg~\cite{Frixione:2007vw} r.\ 1556 and \pythiapp v.\ 8.170 along with the AU2 tune and the CT10~\cite{Lai:2010vv} PDF set and used only for calibration and studies of systematic uncertainties.

\section{ATLAS detector}

The ATLAS detector~\cite{Aad:2008zzm} is a multi-purpose particle detector with a forward-backward symmetric cylindrical geometry and near $4\pi$ coverage in solid angle.\footnote{ATLAS uses a right-handed coordinate system with its origin at the nominal interaction point in the centre of the detector and the $z$-axis coinciding with the axis of the beam pipe. The $x$-axis points from the interaction point to the centre of the LHC ring, and the $y$-axis points upward. Cylindrical coordinates ($r$, $\phi$) are used in the transverse plane, $\phi$ being the azimuthal angle around the beam pipe. The pseudorapidity is defined in terms of the polar angle $\theta$ as $\eta = - \ln \tan(\theta/2)$.} The search for heavy long-lived charged particles relies on measurements of ionisation and time-of-flight, therefore the detector components providing these observables are described below.

\subsection{Pixel detector} \label{sec:pixel}

As the innermost detector system in ATLAS, the silicon pixel detector typically provides at least three high-precision spatial measurements for each track in the region $|\eta|<2.5$ at radial distances from the LHC beam line of $r<15$~cm. The sensors in the pixel barrel ($|\eta|<2$) are placed on three concentric cylinders around the beam-line, whereas sensors in the end-cap ($|\eta|>2$) are located on three disks perpendicular to the beam axis on each side of the barrel. The data are only read out if the signal is larger than a set threshold.

\subsubsection{Specific ionisation measurement}

The charge collected in each pixel is measured using the time-over-threshold (ToT) technique. The calibration of the ToT to the charge deposition in each pixel is established in dedicated scans, and therefore the ToT measurement yields the energy loss of a charged particle in the pixel detector. 

The maximum ToT value corresponds to 8.5 times the average charge released by a minimum ionising particle (MIP) with a track perpendicular to the silicon detectors and leaving all of its ionisation charge on a single pixel. If this value is exceeded, no hit is registered. 

In LHC collisions the charge generated by a charged particle crossing a layer of the pixel detector is usually contained in a few pixels. Neighbouring pixels are joined together to form clusters and the charge of a cluster is calculated by summing the charges of all pixels after calibration correction. The specific energy loss (\dedx) is measured using the average of all individual cluster charge measurements for the clusters associated with the track, typically three measurements. To reduce the effect of tails in the expected Landau distribution, the average is evaluated after removing the cluster with the highest charge (the two clusters with the highest charge are removed for tracks having five or more clusters).

\subsubsection{Mass measurement}

The masses of slow charged particles can be measured using the ID information by evaluating a function that parameterises the expected behaviour of the specific energy loss as a function of the particle $\beta\gamma$. The parametric function describing the relationship between the most probable value of the specific energy loss (${\cal{MPV}}_{\mathrm{d}E\over \mathrm{d}x}$) and $\beta\gamma$ was found by searching for a functional form which adequately describes the simulated data~\cite{dedxnote}. ${\cal{MPV}}_{\mathrm{d}E\over \mathrm{d}x}$ is described via five fixed parameters $p_1$--$p_5$, evaluated separately for data and MC simulation, using

\vspace{-1em}
\begin{equation}
{\cal{MPV}}_{\mathrm{d}E\over \mathrm{d}x}(\beta\gamma) = {p_1\over \beta^{p_3}}\ln(1+(|p_2|\beta\gamma)^{p_5})-p_4.
  \label{bbfun}
\end{equation}

The most probable value of \dedx for MIPs is about 1.2~\MeVgcm with a spread of about 0.2~\MeVgcm and a slight $\eta$ dependence, increasing by about 10\% from low-$|\eta|$ to high-$|\eta|$ regions~\cite{Aad:2011hz}.  The measurable $\beta\gamma$ range lies between 0.2 and 1.5, the lower bound being defined by the overflow in the ToT spectrum, and the upper bound by the overlapping distributions in the relativistic-rise branch of the curve. 

A mass estimate $m_{\beta\gamma} = p/\beta\gamma$ can be obtained for all tracks with a measured specific energy loss \dedx above the value for MIPs,  using their reconstructed momentum $p$ and $\beta\gamma$ evaluated from \dedx. The stability of the measurement of the specific energy loss as a function of time is monitored through measurements of the masses of kaons and protons with percent-level precision and is found to have a variation of less than one percent. For LLPs considered in this article the expected \dedx values can be significantly larger than those of SM particles, allowing their identification based on this information. The RMS of the $m_{\beta\gamma}$ distribution obtained in this way is about 20\%.

\subsection{Calorimeters} \label{sec:calorimeters}

Liquid argon is used as the active detector medium in the electromagnetic (EM) barrel and end-cap calorimeters, as well as in the hadronic end-cap (HEC) calorimeter. All are sampling calorimeters, using lead plates as absorbers material for the EM calorimeters and copper plates as absorbers material for the HEC calorimeter. The barrel EM calorimeter covers the region $|\eta|<1.475$ and consists of a pre-sampler and three layers at radii from 150 to 197~cm. The EM end-cap calorimeter consists of three layers in the region $1.375<|\eta|<2.5$ (two for $2.5<|\eta|<3.2$) and a pre-sampler for $1.5<|\eta|<1.8$. The four layers of the HEC calorimeter cover the range $1.5<|\eta|<3.2$. The time of the energy deposition in each element of the calorimeter (cell) is measured. The typical cell time resolution is 1.5--2.0~ns for energy deposits of 1~\GeV in the EM, 2.0--2.5~ns for energy deposits of 10~\GeV in the HEC.

The ATLAS tile calorimeter is a cylindrical hadronic sampling calorimeter. It uses steel as the absorber material and plastic scintillators as the active material. It covers radii from 228 to 423~cm. The calorimeter is subdivided into a central barrel covering $|\eta| \lesssim 1.0$ and extended barrels covering $0.8 \le |\eta| \le 1.7$. Each barrel part is divided into 64 modules in $\phi$ and the cells in each module are divided into three layers. The typical cell time resolution is 0.6--0.8~ns for energy deposits of 1~\GeV. The time resolution is approximately proportional to $E^{-1/2}$.

\subsection{Muon system} \label{sec:muondetectors}

The muon spectrometer forms the outer part of the ATLAS detector, detects charged particles exiting the calorimeters and measures their momenta in the pseudorapidity range $|\eta| < 2.7$. It is also designed to trigger on these particles in the region $|\eta| < 2.4$. In the barrel the chambers are arranged in three concentric cylindrical shells around the beam axis with radii of 5 to 10~m, while in the two end-caps the muon chambers are arranged in three wheels that are perpendicular to the beam axis at distances between 7.4 and 21.5~m to the nominal interaction point.

The precision momentum measurement is performed by monitored drift tube (MDT) chambers. These chambers consist of three to eight layers of drift tubes covering the region $|\eta| < 2.7$, except in the innermost tracking layer of the forward region ($2.0 < |\eta| < 2.7$), where cathode strip chambers are used. Resistive plate chambers (RPC) in the barrel region ($|\eta|<1.05$) and thin gap chambers (TGC) in the end-cap ($1.05 < |\eta| < 2.4$) provide a fast first-level trigger (level-1). Muons typically have around 20 MDT hits and 10 RPC hits if they traverse the MS barrel, with a typical time resolution of 3.5~ns and 0.6--1.1~ns, respectively. In contrast to the standard ATLAS muon reconstruction, candidate tracks are refitted allowing their velocity to be less than the speed of light, in order to associate all the MDT hits produced by the LLP with the candidate.

\subsection{Measurement of $\beta$ based on time-of-flight} \label{sec:tofbeta}

\begin{figure*}[tb]
  \centering
  \includegraphics[width=0.49\linewidth]{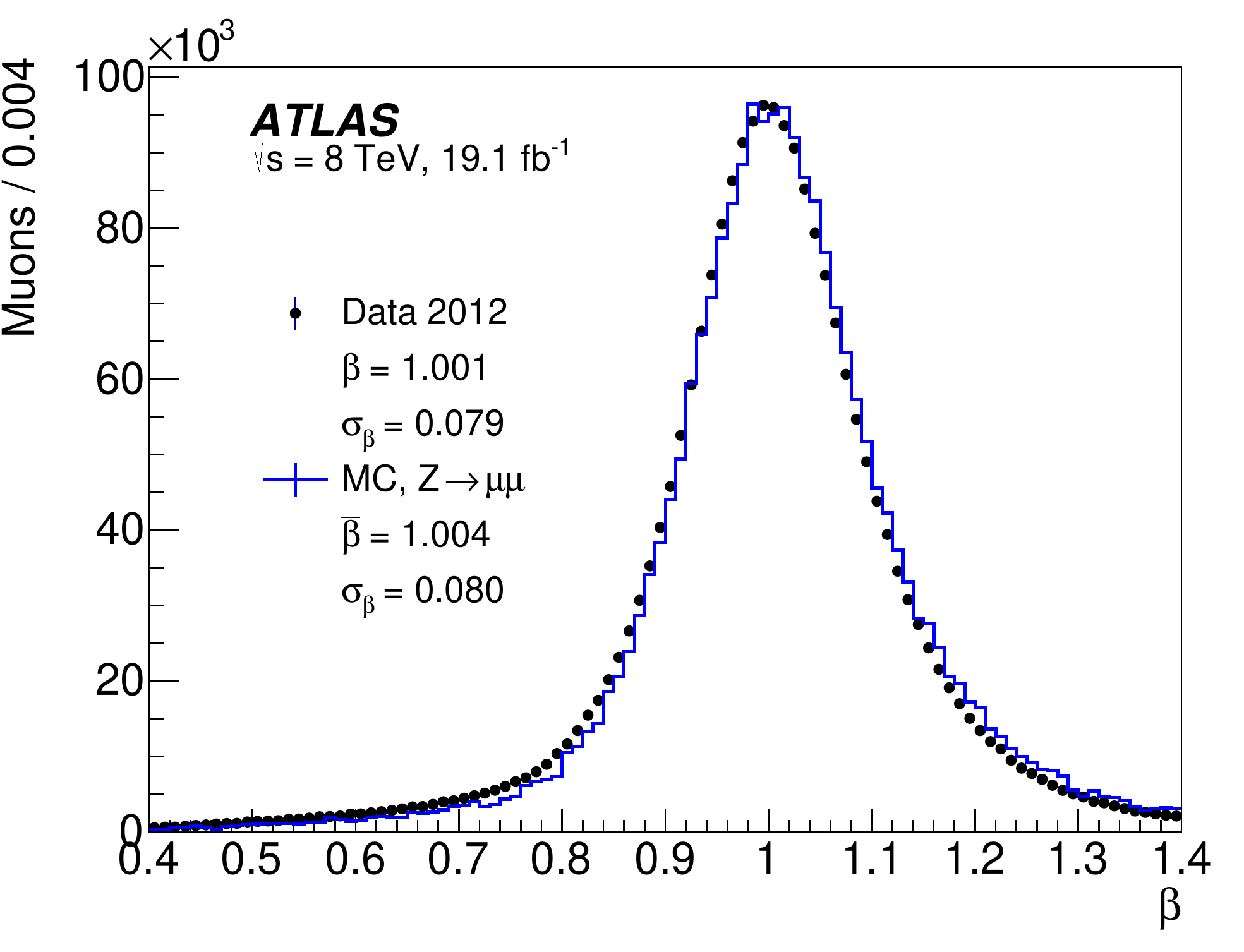}
  \includegraphics[width=0.49\linewidth]{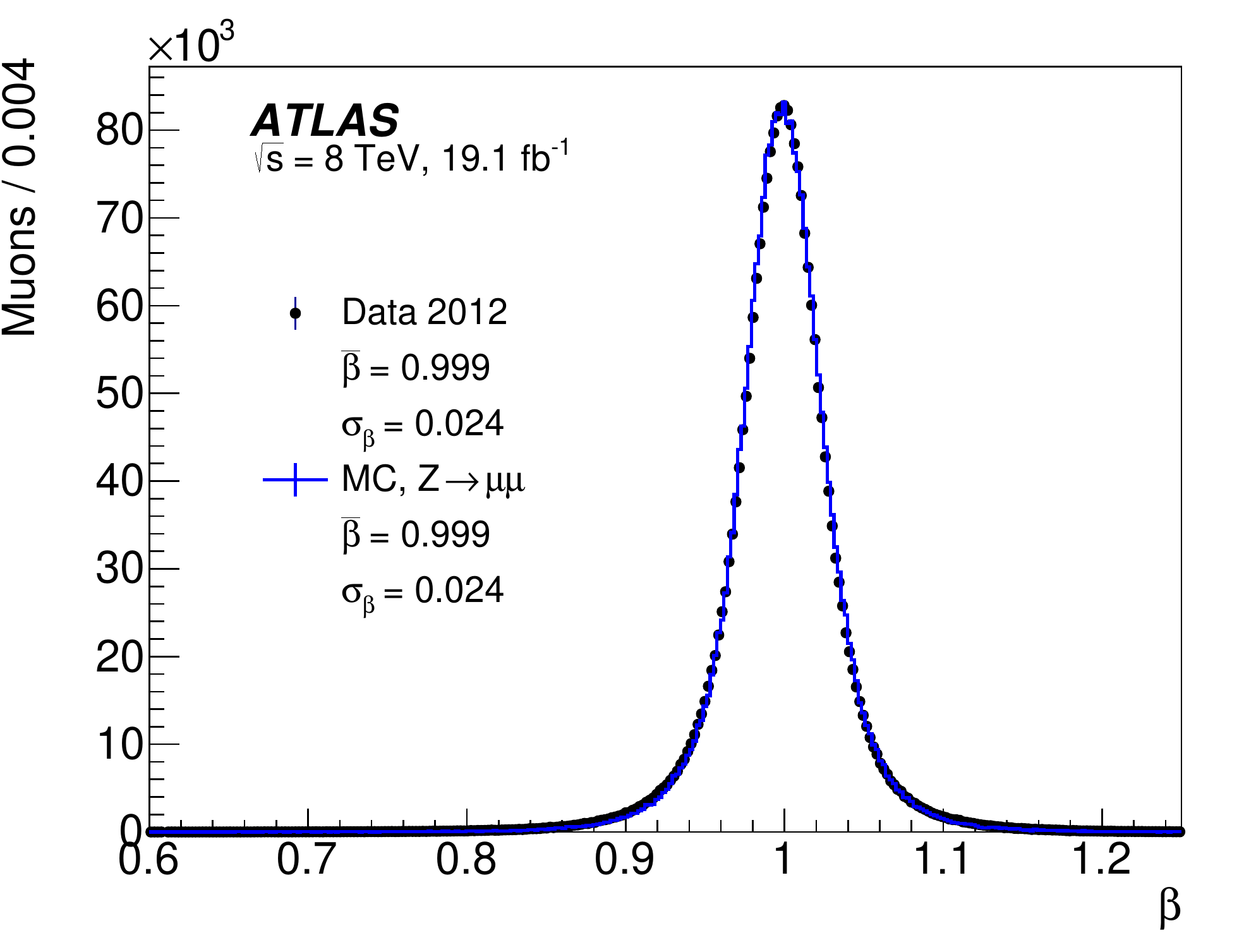}
  \caption{Distribution of the muon speed, $\beta$, from the calorimeter (left) and combined measurements of calorimeter and muon spectrometer (right) obtained for selected \zmumu events in data (points) and smeared MC simulation (line).}
  \label{fig:beta}
\end{figure*}

The calorimeters, RPCs and MDTs have sufficiently accurate timing to distinguish between highly relativistic SM particles and slower LLPs of interest to the searches described in this article. The measured time-of-flight to calorimeter cells and MS hits crossed by the candidate track are used to measure the speed $\beta$. Custom calibration methods using \zmumu events are used to achieve optimal $\beta$ resolution. The times are first corrected collectively for any timing differences between the LHC and ATLAS in order to compensate for collective time-dependent effects (average for each LHC store) and then individually by detector element for any offsets. The calibration also provides a $\beta$ uncertainty for each detector element.
In order to obtain the correct signal efficiency, hit time measurements in simulated \zmumu samples are smeared to correspond to the distribution observed in data. Each hit is smeared by the time resolution observed in the detector element where it was measured.

The individual $\beta$ measurements are combined in a weighted average, using the errors determined per detector element in the calibration. The combination is done first for each sub-detector (calorimeters, MDT, RPC) separately, and then for the entire detector. At each step the measurements are required to be consistent, as described in section~\ref{sec:selection_offline_event_slepton}. Depending on $\eta$, the $\beta$ resolution for muons with $\beta=1$ is 2.4--2.6\% for the RPC, 3.7--4.9\% for the MDT and about 8\% for the calorimeter. Though the calorimeter $\beta$ resolution is less precise than that of the MS, it provides high efficiency and the model independence of the MS-agnostic \rhad search.

Figure~\ref{fig:beta} shows the $\beta$ from the calorimeter (left) and combined measurements (right) obtained for selected \zmumu events in data and smeared MC simulation. The mean values and resolution of the combined $\beta$ are $\bar\beta=0.999$ and $\sigma_\beta=0.024$, respectively, for both data and MC simulation. The RMS of the $m_{\beta}$ distribution obtained this way is about 10\% (20\% for the calorimeter only measurement).

\section{Online event selection} \label{sec:selection_trigger}

All searches are based on events collected by at least one of two trigger types: single-muon and \etmiss triggers. 

\subsection{Single-muon trigger} \label{sec:selection_trigger_muon}

The muon trigger and its performance in 2012 data are described in detail in reference~\cite{Aad:2014sca}. The searches use un-prescaled muon triggers with a transverse momentum (\pT) threshold of 24~\GeV. Offline candidates are selected with $\pT > 70~\GeV$, well above the trigger threshold. 

Events selected by level-1 muon triggers are accepted and passed to the high-level trigger only if assigned to the collision bunch crossing. Late triggers due to the arrival of particles in the next bunch crossing are thus lost. The trigger efficiency for particles arriving late at the MS cannot be assessed from data, where the vast majority of candidates are in-time muons and where low-$\beta$ measurements are due to mismeasurement. The trigger efficiency is thus obtained from simulated signal events. However, the quality of the estimate depends on the accuracy of the timing implementation in the simulation. A detailed emulation of the level-1 electronics circuits, including their timing, is applied to simulated events. The probability that an LLP triggers the event increases roughly linearly from zero at $\beta = 0.62$ to a maximum value of about 70\% at $\beta = 0.82$ for LLPs that reach the MS. A systematic uncertainty is assigned to account for differences in the input time measurements between data and simulated events (see section~\ref{sec:syst_signal}).

GMSB and \lsusy events have two LLPs and possibly muons present in the decay chain, so the likelihood of one of the penetrating particles arriving in the collision bunch crossing is high. Chargino events have no muons, and since in the majority of the chargino events there is only one LLP, the efficiency is lower. The estimated trigger efficiency for GMSB slepton events is between 65\% and 80\%, for \lsusy events between 75\% and 90\% and for stable-chargino events between 24\% and 64\%. Muon triggers are less efficient for \rhads (0--20\%), where one or both of the \rhads may be neutral as they enter the MS and $\beta$ is typically low.

\subsection{Missing transverse momentum trigger} \label{sec:selection_trigger_met}

The \etmiss quantity used at the trigger level is based on the calorimeter only and does not include any corrections for muon-like objects. LLPs deposit very little of their energy in the calorimeter, and therefore in most signal types \etmiss is dominated by initial state radiation (ISR) jets recoiling against the two LLPs. When charginos and neutralinos decay into long-lived sleptons, additional \etmiss may result from neutrinos.

The \rhad searches use un-prescaled \etmiss triggers~\cite{ATLAS-CONF-2014-002} with thresholds as low as 60~\GeV, while the other searches use thresholds between 70 and 80~\GeV. The onset of these triggers is at about 10~\GeV below the threshold, while full efficiency is reached at about 70--80~\GeV above the threshold. Unlike the single-muon trigger, there is no loss of efficiency for the \etmiss triggers when LLPs have low $\beta$.

\subsection{Trigger efficiency} \label{sec:selection_trigger_eff}

In all the searches, except the MS-agnostic \rhad search, a logical \texttt{OR} of the muon and \etmiss triggers described above is used. Depending on the mass of the LLP, the total trigger efficiency is between 80\% and 90\% for GMSB slepton events, between 40\% and 66\% for events with stable charginos, above 95\% for \lsusy events and between 22\% and 35\%, depending on mass and type, for events containing \rhads. In the MS-agnostic \rhad search only the \etmiss triggers are used, with an efficiency between 21\% and 26\%.

\section{Offline event and candidate selection} \label{sec:selection_offline}

Three different signal types are studied: sleptons, charginos and \rhads. An overall trigger and event-quality selection, common to all searches, is applied. Given the different expected interactions with the ATLAS detector, a dedicated selection, containing event-based as well as candidate-based criteria, is optimised and applied for each signal type. An overview of the signal regions can be found in table~\ref{tab:SignalRegions}.

\subsection{Common event and candidate selection} \label{sec:selection_offline_event}

\begin{table*}[tb]
  \footnotesize
  \def\arraystretch{1.0}
  \begin{center}
    {\setlength{\tabcolsep}{0em}\begin{tabular*}{\textwidth}{@{\extracolsep{\fill}}llccccccc}
      \hline
      \hline
      Search    &Signal  &LLP mass   &$N_\mathrm{cand}$ &Momentum           &$|\eta|$     &\etmiss      &$\beta$         &$\beta\gamma$ \\
                &regions &[GeV]      &                 &[GeV]              &             &[GeV]        &                &              \\
      \hline
      Sleptons  &\srslcc &175--510   &2                &$\pt > 70$         &$<2.5$       &             &$<0.95$         &consistency   \\
                &\srslc  &175--510   &1                &$\pt > 70$         &$<2.5$       &             &$<0.85$         &consistency   \\
      Charginos &\srchcc &100--800   &2                &$\pt > 70$         &$<2.5$       &             &$<0.95$         &consistency   \\
                &\srchcl &100--800   &1                &$\pt > 70$         &$<1.9$       &$>100^{***}$ &$<0.95$         &consistency   \\
                &\srchc  &100--800   &1                &$\pt > 70$         &$<1.9$       &             &$<0.85$         &consistency   \\
      \rhads    &\srrhma &400--1700  &$\ge$1           &$p > 140$--$200^*$ &$<1.65$      &             &$<0.88$--$0.74$ &$<2.3$--1.15  \\
                &\srrhfd &400--1700  &$\ge$1           &$p > 140$--$200^*$ &$<1.65^{**}$ &             &$<0.88$--$0.74$ &$<2.3$--1.15  \\
      \hline
      \multicolumn{4}{c}{\smaller \smaller $\phantom{ }^* \Delta R_{\mathrm{jet}, \pt > 40 \GeV} > 0.3$, $\Delta R_{\mathrm{track}, \pt > 10 \GeV} > 0.25$}
      & \multicolumn{3}{c}{\smaller \smaller $\phantom{ }^{**}$ only for \idcalo candidates}
      & \multicolumn{2}{c}{\smaller \smaller $\phantom{ }^{***} \Delta \phi_{\mathrm{LLP},\etmiss} > 1.0$}\\
      \hline
      \hline
    \end{tabular*}}
    \caption{Overview of signal regions (SRs), the covered mass range and selection requirements for different types of long-lived particles. The signal regions for the same search are mutually exclusive and combined in the limit setting, except for two \rhad SRs, which each probe a different hypothesis for the particle interactions with the detector. $N_\mathrm{cand}$ denotes the number of LLP candidates considered in the given SR. The $\beta$ and $\beta\gamma$ requirements listed for the \rhad SRs are due to their mass dependence. In addition, all selections have cosmic-ray muon and $Z$ vetoes. For sleptons and charginos $\beta\gamma$ is only used to check for consistency with $\beta$ by requiring $\beta(\mathrm{TOF})-\beta(\dedx)<5\sigma$.\label{tab:SignalRegions}}
  \end{center}
\end{table*}

Collision events are selected by requiring a good primary vertex. Vertices are reconstructed requiring at least three tracks reconstructed in the ID and consistent with the beam spot. The primary vertex is defined as the one with the highest $\sum p^2_T$ of associated tracks.

Since the background increases significantly at high $|\eta|$, due to the large momenta of candidates and decreased ID momentum resolution, those regions are excluded from the definition of signal regions where they may add large backgrounds and where the signal is expected to be more centrally produced (high masses). High-$|\eta|$ regions are considered for selections where other stringent requirements reduce the background, such as two muons or LLP candidates in the event and/or a precise $\beta$ measurement.

Different requirements on \pt and $p$ are placed in the various searches, as \pt is more suitable to suppress very boosted SM background in cases including the high-$|\eta|$ regions, while the the use of $p$ is advantageous in searches focussing on low-$|\eta|$ regions (e.g.\ \rhads), as it is more closely related to mass.

Additional requirements on $\beta$ ($\beta\gamma$) and $m_{\beta}$ ($m_{\beta\gamma}$) are used to reduce background. The presence of an LLP signal is searched for in a distribution of $m_{\beta}$ ($m_{\beta\gamma}$), where the signal should peak and background be continuous.

\subsection{Slepton event and candidate selection} \label{sec:selection_offline_event_slepton}

In GMSB and \lsusy events, the weak coupling of the gravitino to the other particles implies that only the NLSP ($\stau_1$ for the models of interest) decays to the gravitino. As a result of this and of $R$-parity conservation, at least two $\stau_1$ sleptons are expected in each GMSB event, both with a high probability of being observed. Therefore the slepton searches require at least two loosely identified muon-like objects reconstructed using the techniques described in reference~\cite{Aad:2014rra}, which will be called LLP candidates in the following. By applying this selection criterion, background from $W$ and multi-jet events is reduced. Two sets of selection criteria are applied on a per-candidate basis with details given below. A loose selection with high efficiency is used to select candidates in events where there are two LLP candidates, since a background event would very rarely have two high-\pt~muons, both with poorly measured $\beta$ and a large reconstructed mass. Events having two loose candidates, independent of their charge, fall in the two-candidate signal region (\srslcc). In events where only one candidate passes the loose selection, that candidate is required to pass an additional, tighter selection. Such events are collected in a mutually exclusive one-candidate signal region (\srslc).

Candidates in the loose slepton selection are required to have $\pt > 70~\GeV$ and $|\eta| < 2.5$. Any two candidates that combine to give an invariant mass within $10~\GeV$ of the $Z$ boson mass are both rejected. Candidates are also required to have associated hits in at least two of the three layers of precision measurement chambers in the MS. Cosmic-ray muons are rejected by a topological requirement on the combination of any two candidates with opposite $\eta$ and $\phi$. The number of degrees of freedom in the $\beta$ measurement\footnote{The number of calorimeter cells plus MS hits contributing to the $\beta$ measurement minus the number of detector systems} is required to be larger than three. LLP candidates are expected to have low $\beta$ values and these values are expected to be consistent between individual measurements, both in the same detector system and between different detectors, while in the case of muons a low $\beta$ value would be due to a poor measurement in only one of the detectors.
The different detector system measurements of $\beta$ are required to be pair-wise consistent at the 3$\sigma$ level, and the combined $\beta$ to be consistent with the $\beta\gamma$ estimated in the pixel detector within 5$\sigma$. The $\beta$ resolution is estimated for each candidate, and the $\beta\gamma$ resolution is about 11\%. The $\beta\gamma$ measurement is translated to $\beta$ and compared to the value of $\beta$ based on time-of-flight for the consistency check. There is no requirement on the value of $\beta\gamma$ obtained from the pixel \dedx measurement in the searches that require consistency. As a result, many candidates are in the MIP region. Those are required to have $\beta$ consistent with the MIP hypothesis. Finally, in order to reduce the muon background, the combined $\beta$ measurement is required to be between 0.2 and 0.95.

To pass the tighter slepton selection used for \srslc, a candidate is additionally required to have at least two separate detector systems measuring $\beta$ and the number of degrees of freedom of the $\beta$ measurement is required to be at least six.

Events with one candidate are then divided between \srslc, where the combined $\beta$ measurement is required to be less than 0.85, and a control region with $0.85<\beta<0.95$, used to cross-check the background estimation.

Finally, the measured mass, $m_{\beta}=p/\beta\gamma$, calculated from the candidate's momentum and its measured $\beta$, is required to be above some value. The value is chosen according to the mass of the the hypothetical $\stau_1$ mass in the given model, so as to achieve 99\% signal efficiency with respect to the earlier selection. For \srslcc, both masses are required to be above the chosen value.

Typical efficiencies for signal events to satisfy all criteria including the mass requirement are 30\% for \srslcc and 20\% for \srslc, giving 50\% efficiency in total. The efficiencies are similar for events with pair-produced sleptons and events where sleptons arise from directly produced, decaying charginos and neutralinos.

\subsection{Chargino event and candidate selection} \label{sec:selection_offline_event_chargino}

Except for the two-muon requirement, the chargino event selection is the same as the slepton selection. For chargino-pair production, the events would be very similar to slepton events, while for chargino--neutralino production a single LLP candidate is accompanied by \etmiss caused by the neutralino. The chargino and neutralino are typically well-separated in $\phi$, therefore the \ptmiss is expected to point in the opposite direction to the reconstructed LLP. The events are divided into three signal regions. Events with two LLP candidates passing the loose selection, as before independent of their charge, are in the two-candidate signal region (\srchcc). This selection is motivated by pair production of charginos. Events with one candidate passing the loose selection must have \etmiss$>$100~\GeV and an azimuthal angular distance between the LLP candidate and the \ptmiss $\Delta\phi>1$, to be included in the one-loose-candidate signal region (\srchcl). \srchcl is motivated by the chargino--neutralino production mode. Finally if an event has neither two candidates nor large \etmiss, one given candidate has to pass the tighter selection and have $\beta<0.85$ to be included in the one-candidate signal region (\srchc). All three signal regions are mutually exclusive. 

The requirements for a candidate to pass the loose or the tight selection are the same as for the slepton search. In addition, in both \srchcl and \srchc, candidates with $|{\eta}|>1.9$ are excluded.

A mass selection, chosen to achieve 99\% signal efficiency with respect the earlier selection, is applied to the candidate mass. This requirement depends on the hypothetical chargino mass and differs by model. For \srchcc, both masses are required to be above the chosen value.

Typical efficiencies for signal events to satisfy all selection criteria including the mass requirement are 5--6\% for \srchcc, 10--13\% for \srchcl and 3\% for \srchc, giving 18--22\% efficiency in total, depending on the mass of the chargino candidates. Looking separately at the two different production modes, the efficiency of \srchcc for chargino-pair production is 15--20\% and the efficiency of \srchcl for chargino--neutralino production is 12--17\%.

\subsection{\rhad event and candidate selection} \label{sec:selection_offline_event_rhadron}

Since the \rhad contains light quarks and gluons in addition to the squark or gluino, the charge of the \rhad can change following nuclear interactions with the detector material. This possibility makes it difficult to rely on a single detection mechanism without any loss of detection efficiency, as a neutral state would not be detected until the next nuclear interaction occurs. Some of the main hadronic states resulting from such charge exchange in the models considered are neutral. In a search for \rhads that are produced charged, it is therefore natural to take an inside-out approach, starting from the ID track and adding discriminators from outer detector systems, in case a signal is seen along the extrapolated track. This is reflected in the two different \rhad approaches.

In an \idcalo selection, candidates are required to have a good-quality ID track with $\pt > 50~\GeV$ and $|\eta| < 1.65$. To ensure reliable estimates of $\beta\gamma$ and $\beta$, candidates must not be within an $\eta$--$\phi$ distance $\Delta R = \sqrt{\left( \Delta \eta \right)^2 + \left(\Delta \phi \right)^2} = 0.3$ of any jet with $\pt > 40~\GeV$, reconstructed from calorimeter energy clusters using the anti-$k_t$ jet algorithm~\cite{jetalgo} with distance parameter set to 0.4. Furthermore, candidates must not have any nearby ($\Delta R < 0.25$) tracks with $\pt > 10~\GeV$ nor have pixel hits shared with other tracks. The $Z$ boson mass window and cosmic-ray muon rejection are applied in the same way as in the slepton searches. Candidates must have a good \dedx measurement and a good estimate of $\beta$. The uncertainty on the calorimeter-only $\beta$ is required to be less than 12\%.

In a \combined selection, candidates are required to have a combined track, reconstructed in both the ID and the MS. With the exception of the explicit $\eta$ requirement, the ID requirements for the \combined candidate as well as the $Z$ boson mass window, cosmic-ray muon rejection and \dedx measurement are the same as for the \idcalo selection. The estimate of $\beta$, based on a combination of internally consistent measurements in the calorimeter, the RPCs and the MDTs, is required to have an uncertainty of less than 5\%.

In the full-detector \rhad search, candidates are first checked for compatibility with the \combined selection and only when failing, for compatibility with the \idcalo selection. The two types of candidates are therefore mutually exclusive and events containing at least one candidate fulfilling either of the two selections are considered in the full-detector signal region (\srrhfd).

The independent MS-agnostic \rhad search, ignoring MS information, as well as the muon trigger, considers events containing at least one candidate passing the \idcalo selection (\srrhma).

In the approximately 15\% of events with more than one candidate, a candidate passing the combined selection is preferred; if there are two or more candidates from the same category, one is chosen at random and the others are discarded. In both \rhad searches, additional requirements on a minimum momentum and maximum values for $\beta$ and $\beta\gamma$ are set, depending on the mass hypothesis in question. The two \rhad mass estimates $m_{\beta\gamma}$ and $m_{\beta}$ are both required to be larger than the mass-peak value for the given hypothesis minus twice the width of the mass peak, which is typically around 20\% of the peak mass, leading to an efficiency of more than 95\%. All mass and momentum requirements are the same for gluinos, sbottoms and stops, while the requirements on $\beta\gamma$ and $\beta$ are optimised separately to account for the lower expected cross-section in the sbottom and stop cases. The signal efficiency for gluino, sbottom and stop \rhads is typically 8--12\%, 5--9\% and 8--13\%, respectively, in the MS-agnostic search and 8--15\%, 8--11\% and 15--18\%, respectively, in the full-detector search, depending on the mass hypothesis. While stops and sbottoms have the same cross-section, sbottoms tend to hadronise into neutral states (57\%) slightly more often than stops (43\%). In addition, more sbottom-based \rhads convert into neutral states, as they traverse material, than stop-based \rhads do, reducing the efficiency of the sbottom search compared to the stop one.

\section{Background estimation} \label{sec:background}

The background for all searches is almost entirely composed of high-\pt muons with mismeasured $\beta$ and/or large ionisation. Most of this instrumental background is rejected by requiring a $\beta$ measurement significantly smaller than one and by requiring consistency between the different, independent $\beta$ and $\beta\gamma$ measurements. The background estimate is derived from data in all cases. The background mass distribution can be estimated by producing random pairings of momentum and $\beta$ (and $\beta\gamma$ where applicable) according to the distributions seen in the data. The procedure relies on two validated assumptions: that the signal-to-background ratio before applying selections on $\beta$ ($\beta\gamma$) is small, and that the $\beta$ ($\beta\gamma$) distribution for background candidates is due to measurement resolution and is therefore independent of the source of the candidate and its momentum.

To avoid $\beta$--momentum measurement correlations arising from different detector systems and for different pseudorapidity regions, the detector is divided into eight $\eta$ regions so that the $\beta$ resolution within each region is similar.

\subsection{Slepton and chargino searches} \label{sec:background_sleptonchargino}

The muon $\beta$ probability density function (pdf) in each $\eta$ region is the distribution of the measured $\beta$ of muons in the region normalised to unity, and is obtained separately for each signal region from candidates passing the selection described in sections~\ref{sec:selection_offline_event_slepton} and \ref{sec:selection_offline_event_chargino}, but without the requirements on the value of $\beta$ or $m_{\beta}$.

The background is then estimated by drawing a random $\beta$ from the appropriate muon $\beta$ pdf and calculating $m_{\beta}$ using the momentum of the reconstructed LLP candidates only in cases where the $\beta$ satisfies the selection requirement. Events with two candidates before the $\beta$ requirement are used to estimate the background in \srslcc and \srchcc. The statistical uncertainty of the background estimate is reduced by repeating this procedure many times for each candidate and dividing the resulting distribution by the number of repetitions.

\subsection{\rhad searches} \label{sec:background_rhadron}

In the \rhad searches, the pdfs are produced from candidates in data, which satisfy the selection criteria, except those on $\beta$, $\beta\gamma$, $m_\beta$ and $m_{\beta\gamma}$. As each particle/mass hypothesis has a different selection, the background estimates are produced in each case.

The momentum pdf is produced from candidates that pass the momentum requirement, but have $\beta < 0.90$ and $\beta\gamma < 2.5$, while the $\beta$ and $\beta\gamma$ pdfs are produced by selecting candidates which pass the respective $\beta$ and $\beta\gamma$ selection and have momentum in the range $70~\GeV < p < 180~\GeV$. This ensures that enough events are selected for the background pdfs to reflect the signal region even at high masses.
The independence of $p$, $\beta$ and $\beta\gamma$ required for this approach to work is achieved by considering five equidistant regions in $|\eta|$. The typical number of events in the pdfs used for generating the background estimate is $O(10^{4})$.

\section{Systematic uncertainties} \label{sec:syst}

Several possible sources of systematic uncertainty are studied. The resulting systematic uncertainties are summarised in tables~\ref{tab:systematics1} and \ref{tab:systematics2}. The uncertainties given are those on the expected yields in the signal region.

\begin{table*}[tb]
  \footnotesize
  \begin{center}
    {\setlength{\tabcolsep}{0em}\begin{tabular*}{\textwidth}{@{\extracolsep{\fill}}lrrrr}
      \hline\hline
      & \multicolumn{2}{c}{GMSB} & \multicolumn{2}{c}{\lsusy} \\
      Source                              & \srslc & \srslcc & \srslc & \srslcc  \\    
      \hline
      \hline
      Signal size -- theory               & 5         & 5         & 1--54     & 1--54     \\
      \hline
      Signal efficiency                   &           &           &           &           \\
      ~$\cdot$~Trigger efficiency         & 3.2       & 3.2       & 3.1       & 3.1       \\
      ~$\cdot$~ISR                        & $\leq$0.5 & $\leq$0.5 & $\leq$0.5 & $\leq$0.5 \\
      ~$\cdot$~Pixel \dedx calibration    & 1.1       & 1.1       & 1.1       & 1.1       \\
      ~$\cdot$~$\beta$ timing calibration & 1.0       & 2.0       & 1.0       & 2.0       \\
      Total signal efficiency             & 3.6       & 4.0       & 3.5       & 3.9       \\
      \hline
      Luminosity                          & 2.8       & 2.8       & 2.8       & 2.8       \\
      \hline
      Background estimate                 & 10--12    & 8.3--9     & 10--12    & 8.3--9     \\
      \hline
      \hline
    \end{tabular*}}
    \caption{Summary of systematic uncertainties for the slepton searches (given in percent). Ranges indicate a mass dependence for the given uncertainty (low mass to high mass).}
    \label{tab:systematics1}
  \end{center}
\end{table*}

\begin{table*}[tb]
  \footnotesize
  \begin{center}
    {\setlength{\tabcolsep}{0em}\begin{tabular*}{\textwidth}{@{\extracolsep{\fill}}lrrrr}
      \hline\hline
      & \multicolumn{3}{c}{Charginos} & \rhads \\
      Source                              & \srchc & \srchcl & \srchcc  & \srrhma \& \srrhfd \\    
      \hline
      \hline
      Signal size -- theory               & 8.5       & 8.5       & 8.5       & 15--56     \\
      \hline
      Signal efficiency                   &           &           &           &            \\
      ~$\cdot$~Trigger efficiency         & 3.4       & 3.4       & 3.4       & $\leq$2.4  \\
      ~$\cdot$~ISR                        & $\leq$1.0 & $\leq$1.0 & $\leq$1.0 & $\leq$9    \\
      ~$\cdot$~Pixel \dedx calibration    & 1.1       & 1.1       & 1.1       & 1.1        \\
      ~$\cdot$~$\beta$ timing calibration & 1.0       & 1.0       & 2.0       & $\leq$3.6  \\
      ~$\cdot$~Offline \etmiss scale      & 5.6--7.6  & 2--4.2    &           &            \\
      Total signal efficiency             & 6.8--8.5  & 4.3--5.7  & 4.2       & $\leq$10.2 \\
      \hline
      Luminosity                          & 2.8       & 2.8       & 2.8       & 2.8        \\
      \hline
      Background estimate                 & 3.5--6.8  & 4         & 8.7--20   & 3--15      \\
      \hline
      \hline
   \end{tabular*}}
   \caption{Summary of systematic uncertainties for the chargino and \rhad searches (given in percent). Ranges indicate a mass dependence for the given uncertainty (low mass to high mass).}
   \label{tab:systematics2}
  \end{center}
\end{table*}

\subsection{Theoretical cross-sections} \label{sec:syst_theo}

Signal cross-sections are calculated to next-to-leading order in the strong coupling constant, including the resummation of soft gluon emission at next-to-leading-logarithm accuracy (NLO+NLL)\footnote{The NLL correction is used only for strong squark and gluino production when the squark and gluino masses lie between 200~\GeV and 2~\TeV. Following the convention used in the NLO calculators the squark mass is defined as the average of the squark masses in the first two generations. In the case of gluino-pair (associated squark--gluino) production processes, the NLL calculations are extended up to squark masses of 4.5~\TeV (3.5~\TeV). For masses outside this range and for other types of production processes (i.e. electroweak and associated strong and electroweak), cross-sections at NLO accuracy obtained with \prospino~\cite{Beenakker:1996ch} are used.}~\cite{Beenakker:1997ut, Beenakker:2010nq, Beenakker:2011fu}. The nominal cross-section and the uncertainty are taken from an envelope of cross-section predictions using different parton distribution function sets and factorisation and renormalisation scales, as described in reference~\cite{Kramer:2012bx}.

The procedure results in an uncertainty of 5\% for the GMSB slepton search (dominated by electroweak production), between 1\% (low squark mass) and 54\% (high squark mass) in the \lsusy slepton search, 8.5\% in the stable-chargino search and from 15\% (at 400~\GeV) to 56\% (at 1700~\GeV) in the \rhad searches.

\subsection{Signal efficiency} \label{sec:syst_signal}

The muon trigger efficiency for muons is calculated using the tag-and-probe technique on \zmumu events as described in reference~\cite{Aad:2014sca}. The reduction in the muon trigger efficiency due to late arrival of particles is estimated from simulation. However, the quality of the estimate depends on the exact timing implementation in the simulation, and needs to agree well with the data in order to obtain a good estimate of the trigger efficiency. For events triggered by the RPC, a systematic uncertainty is estimated by smearing the hit times in the simulation according to uncalibrated data, and applying the trigger efficiency as a function of $\beta$. The uncertainty on events triggered by the TGC is negligible due to an accurate timing description in the simulation. The resulting systematic uncertainty on the muon trigger efficiency is between 2.9\% and 3.4\%, depending on the signal model.

The \etmiss triggers use calorimeter energy deposits to calculate the transverse energy, and are thus blind to muons. Therefore, \zmumu events can be used for calibration and to study systematic uncertainties. To evaluate the trigger efficiency, the trigger turn-on curve is obtained by fitting the measured efficiency as a function of \etmiss in \zmumu events, both in data and simulation. These efficiency turn-on curves are then applied to the expected \etmiss spectrum from simulated signal events. The total uncertainty is estimated from three contributions: the relative difference between the efficiencies obtained using the fitted threshold curves from \zmumu data and simulation\hide{ \raisebox{.5pt}{\textcircled{\raisebox{-.1pt} {\scriptsize 1}}}} as well as the differences in efficiency obtained from independent $\pm 1\sigma$ variations in fit parameters relative to the unchanged turn-on curve fit for both \zmumu data\hide{ \raisebox{.5pt}{\textcircled{\raisebox{-.1pt} {\scriptsize 2}}}} and MC simulation\hide{ \raisebox{.5pt}{\textcircled{\raisebox{-.1pt} {\scriptsize 3}}}}. The total estimated \etmiss trigger uncertainties are 1.2\%, 3.4\%, 2.5\%, 2.3\% and 2.3\% for sleptons, stable charginos as well as gluino, sbottom and stop \rhads, respectively. These uncertainties include effects of a 10\% variation of the \etmiss scale, motivated by comparing the calorimeter response in \zmumu events between data and simulation.

The trigger efficiency depends on the amount of ISR. To evaluate the associated uncertainty, a number of representative mass hypotheses are reproduced, setting the \pythia radiation level low and high~\cite[p.~391]{Sjostrand:2006za}. A simple threshold curve modelling of the trigger is applied to those and the nominal samples. The largest variation from the central sample is found to be between 0.5\% and 1\% for the slepton and chargino searches and below 9\% for the \rhad searches.

The systematic uncertainty on the efficiency of the offline \etmiss selection for \srchcl is determined by varying the energy or momentum scale of the individual components entering the calculation, and propagating these changes to the \etmiss calculation. The dominant contributions are the muon momentum scale (at low masses) and the jet energy scale~\cite{Aad:2014bia} (at high masses). A systematic uncertainty on the signal efficiency can be obtained by using those scaled values when applying the \etmiss requirement and is estimated to be between 2\% and 7.6\%. \srchc is affected by candidates migrating in or out of \srchcl.

The signal $\beta$ resolution is estimated by smearing the measured time of hits in the MS and calorimeter according to the spread observed in the time calibration. The systematic uncertainty due to the smearing process is estimated by scaling the smearing factor up and down, so as to bracket the distribution obtained in data. A 1\% (2\%) systematic uncertainty is found in \srslc and \srchc (\srslcc and \srchcc) for the slepton and chargino searches, respectively. The corresponding uncertainty for \rhads is estimated to be less than 3.6\% (1\%) in the full-detector (MS-agnostic) approach.

The $\beta\gamma$ measurement from the pixel \dedx carries a systematic uncertainty due to the difference between simulation (signal) and data (predominantly background). This difference can be measured using \zmumu events in data and simulation, and the scale between data and MC simulation is found to be different by 2.3\%. In addition, variations in the $\beta\gamma$ scale are checked, assuming tracking to be very stable, by monitoring the measured proton mass over time. This results in an RMS of 0.6\%, yielding a total scale uncertainty of 2.4\%. Applying this scale uncertainty in simulated signal events leads to uncertainties on the signal efficiency of
1.1\% at low \rhad masses (300~\GeV) and 0.4\% at higher masses (500 and 800~\GeV). For simplicity, a systematic uncertainty of 1.1\% is applied for all masses.

The uncertainty on the integrated luminosity is $\pm$2.8\%. It is derived, following the same methodology as that detailed in reference~\cite{Aad:2013ucp}, from a preliminary calibration of the luminosity scale derived from beam-separation scans performed in November 2012.

\subsection{Background estimation} \label{sec:syst_bkg}

To test the momentum dependence of the muon $\beta$ pdf, the candidates in each $\eta$ region are split into a high and a low momentum category with the same amount of events in each, and the background is estimated with the resulting $\beta$ pdfs. The effect on the background estimation due to residual $p$--$\beta$ correlations is assessed by using a finer $|\eta|$ division in the slepton searches. The detector is sub-divided into 25 $|\eta|$ regions instead of the 8 used in the analysis and the background is estimated with this division. 

Similar tests are performed to determine the uncertainty on the background for the \rhad search. Unlike the slepton search, the range of mass hypotheses tested is very large, and it is found that the size of the systematic uncertainty on the background estimate grows with mass. In order to quantify the systematic uncertainty, the pdfs used to produce the background estimates are varied, both by changing the selection range used for producing them and also by dividing the selection ranges into two sub-ranges, and comparing the resulting background estimates from these sub-range pdfs.

The total uncertainties on the background estimate are 8.3--9\% for \srslcc increasing with mass, and 10--12\% for \srslc in the slepton searches. The uncertainties in the chargino searches are 3.5--6.8\% for \srchc, 4\% for \srchcl and 8.7--20\% for \srchcc. For \rhads the uncertainty is 3--15\%.

\section{Results}

\begin{figure*}[p]
  \centering
  \includegraphics[width=0.48\linewidth]{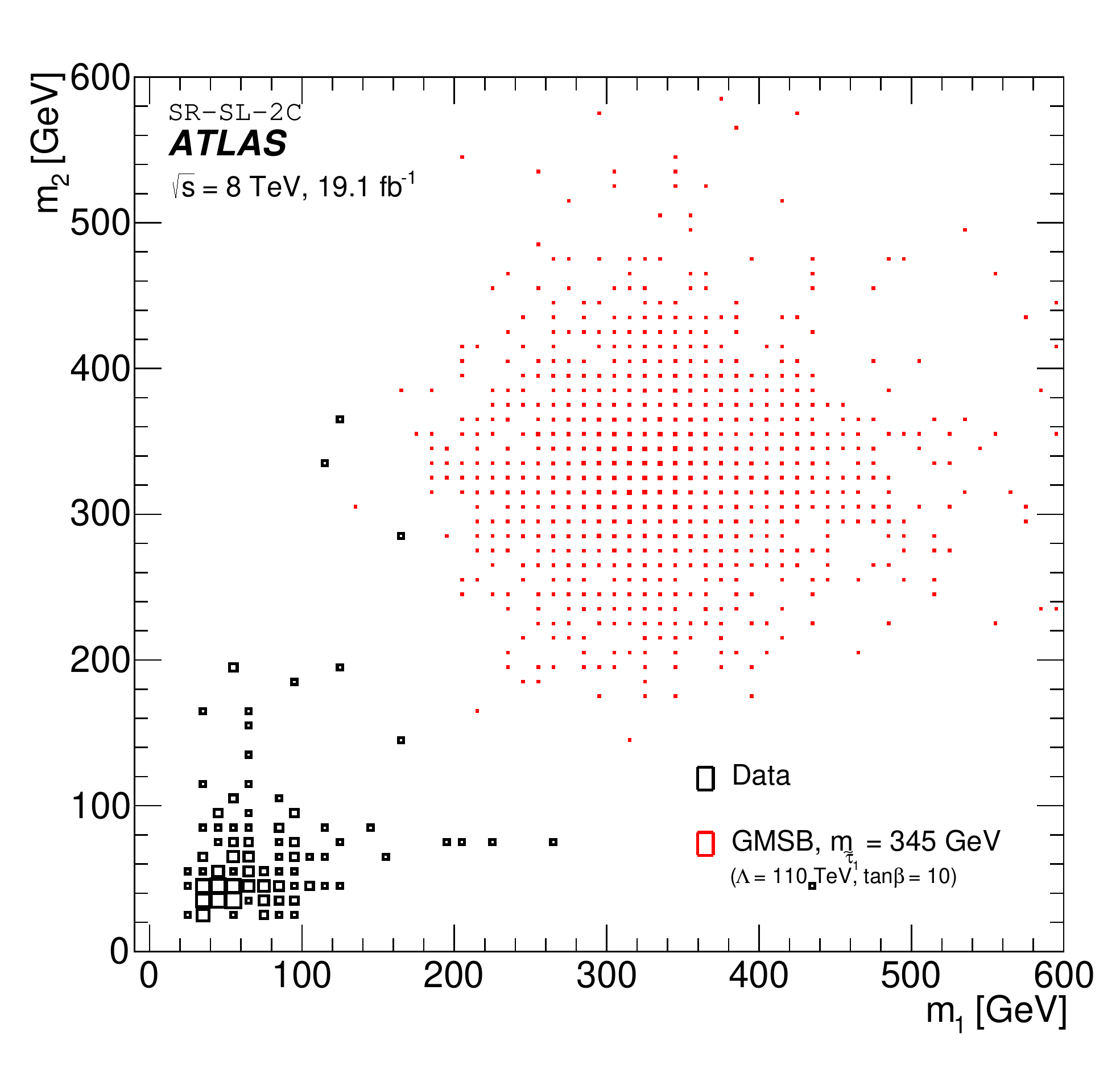}
  \includegraphics[width=0.48\linewidth]{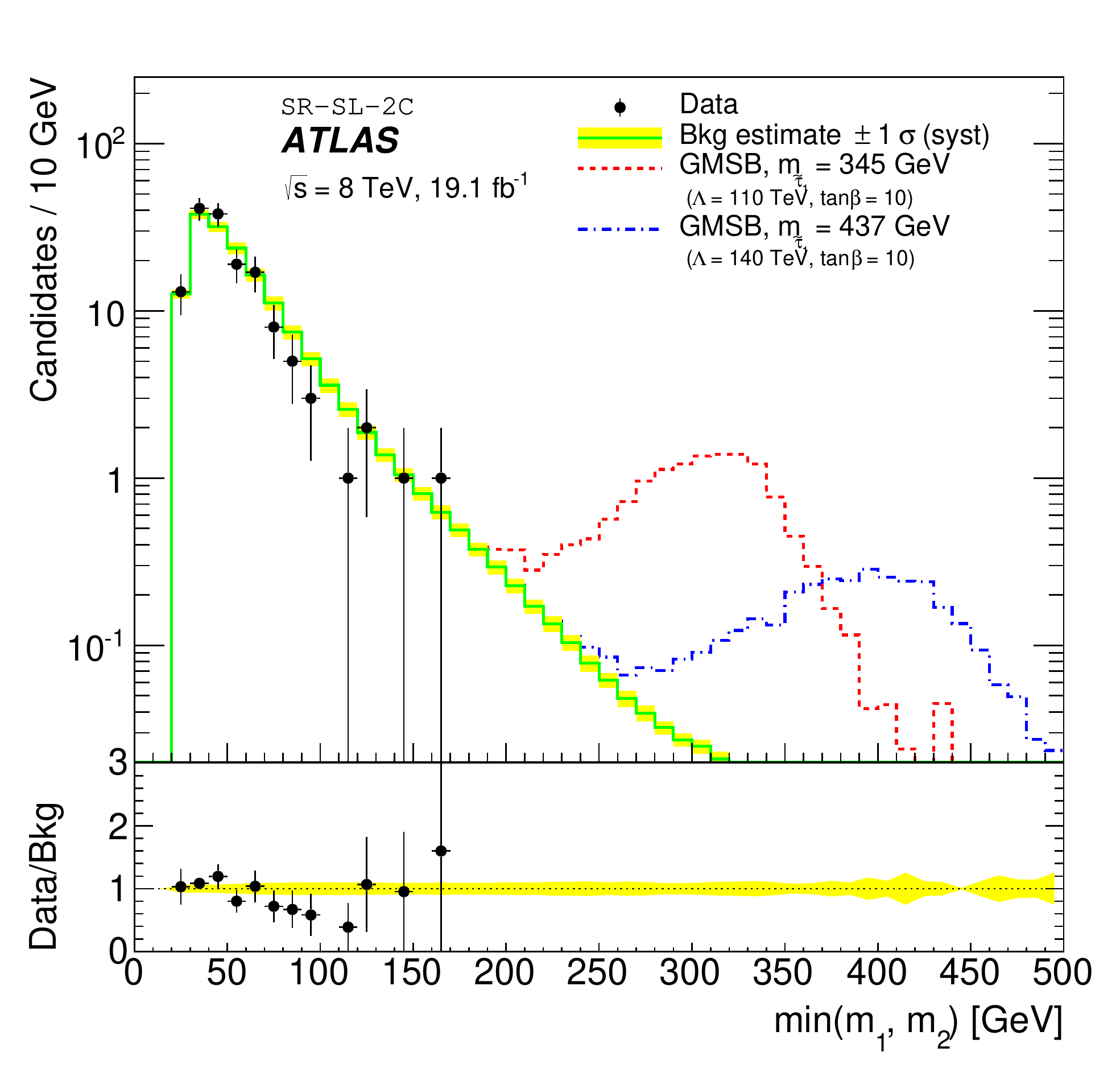}
  \includegraphics[width=0.48\linewidth]{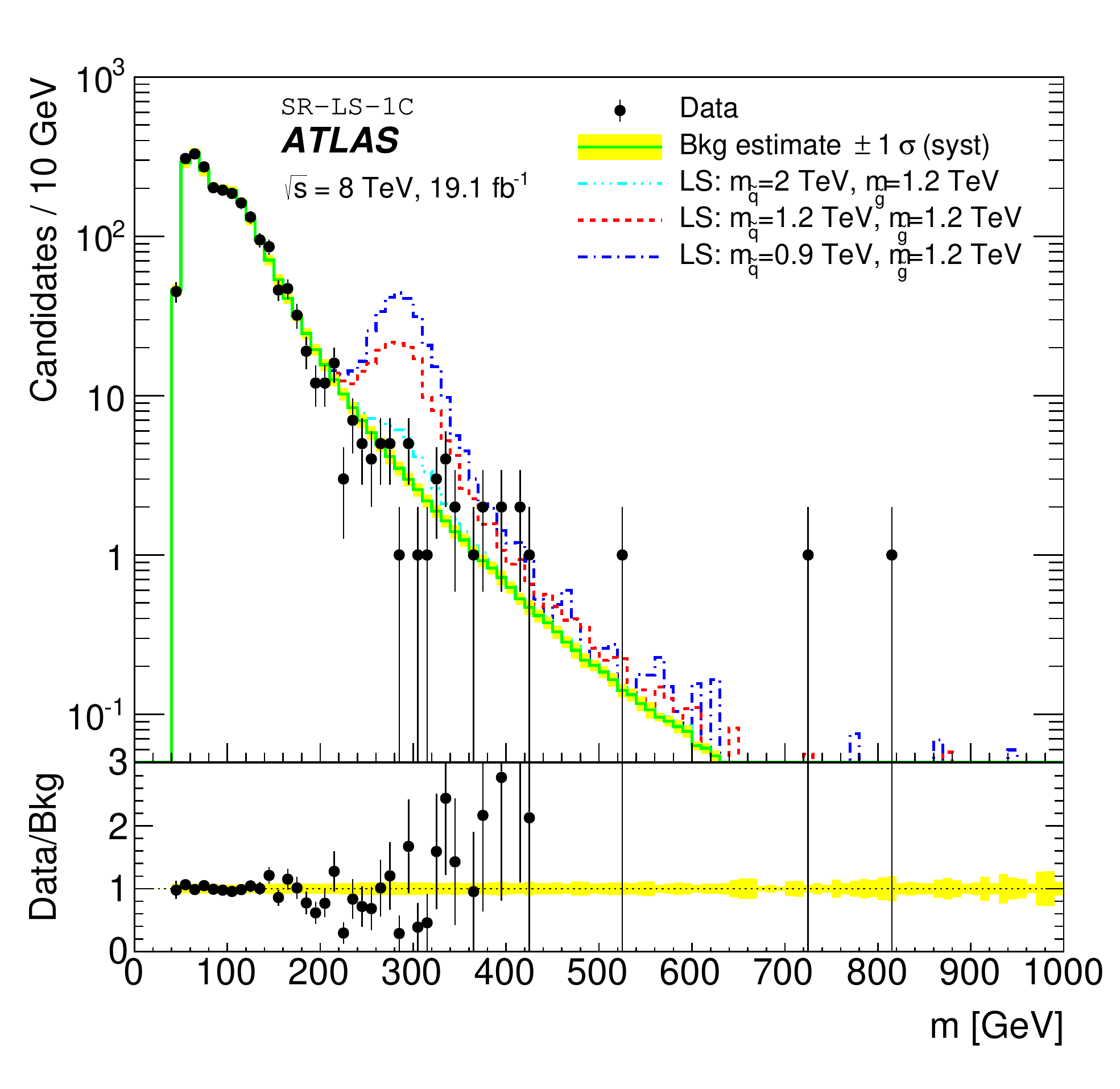}
 \caption{Reconstructed mass $m_\beta$ of one candidate ($m_2$) versus $m_\beta$ of the other candidate ($m_1$) for observed data and expected signal, in the GMSB slepton search in the two-candidate signal region (top-left). Observed data, background estimate and expected signal in the slepton search for the lower of the two masses ($m$) in the two-candidate signal region (GMSB $\stau_1$ masses of 344.5 and 437~\GeV; top-right) and for the one-candidate signal region (\lsusy $m_{\tilde q}$ = 2.0, 1.2 and 0.9~\TeV with $m_{\tilde g}$ = 1.2~\TeV; bottom).} 
  \label{fig:DSB}
\end{figure*}

\begin{figure*}[p]
  \centering
  \includegraphics[width=0.48\linewidth]{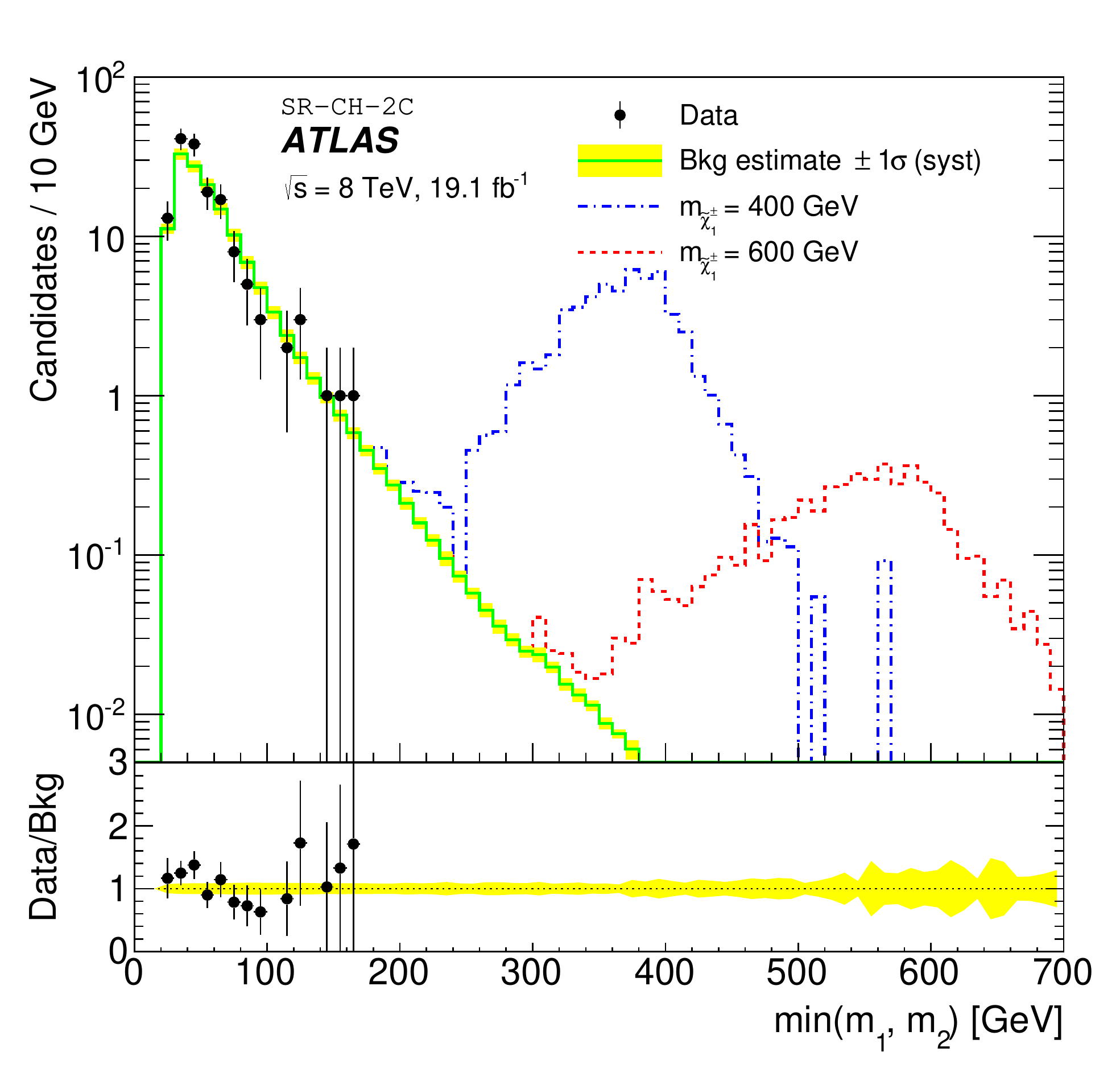}
  \includegraphics[width=0.48\linewidth]{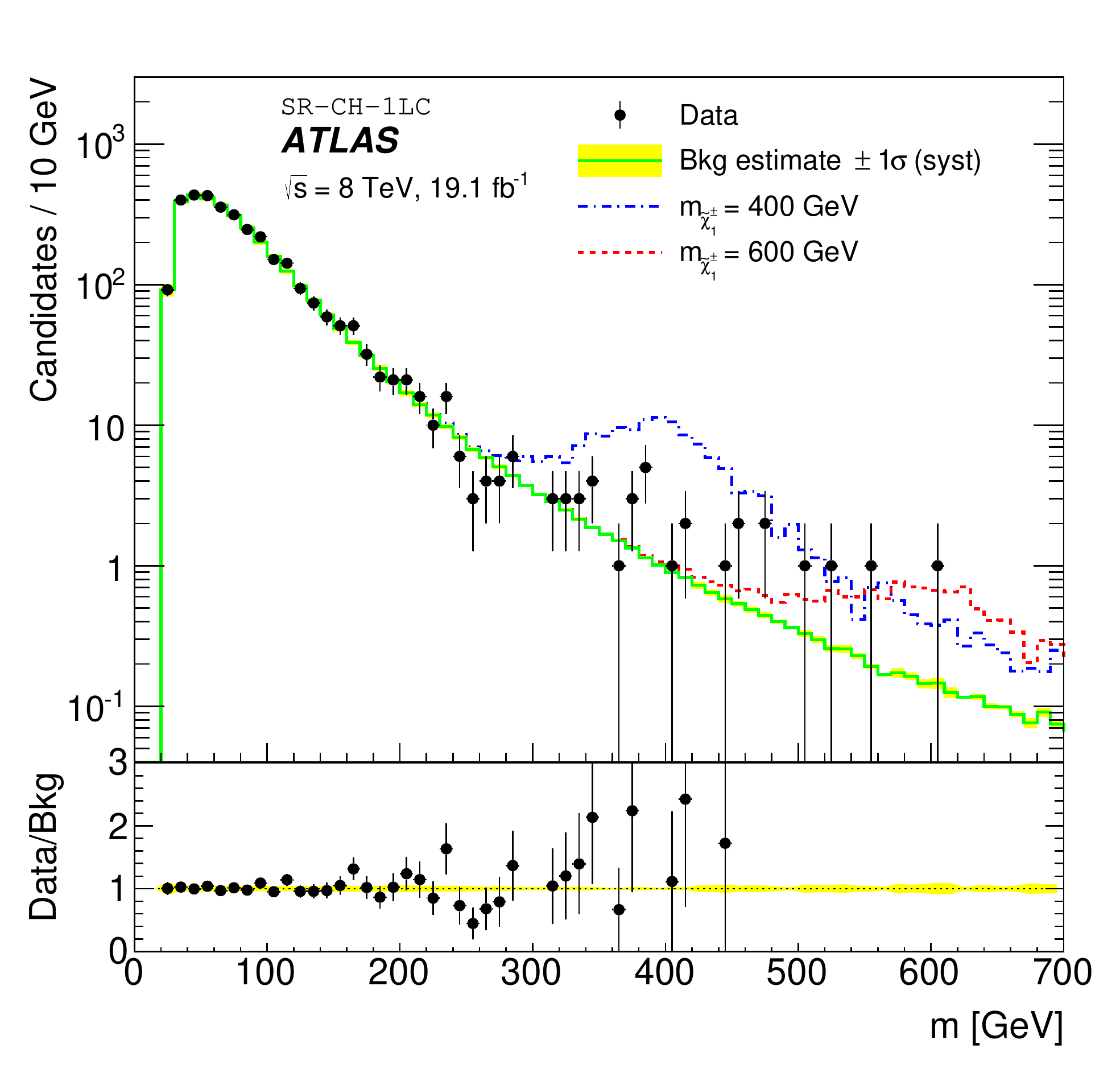}
  \includegraphics[width=0.48\linewidth]{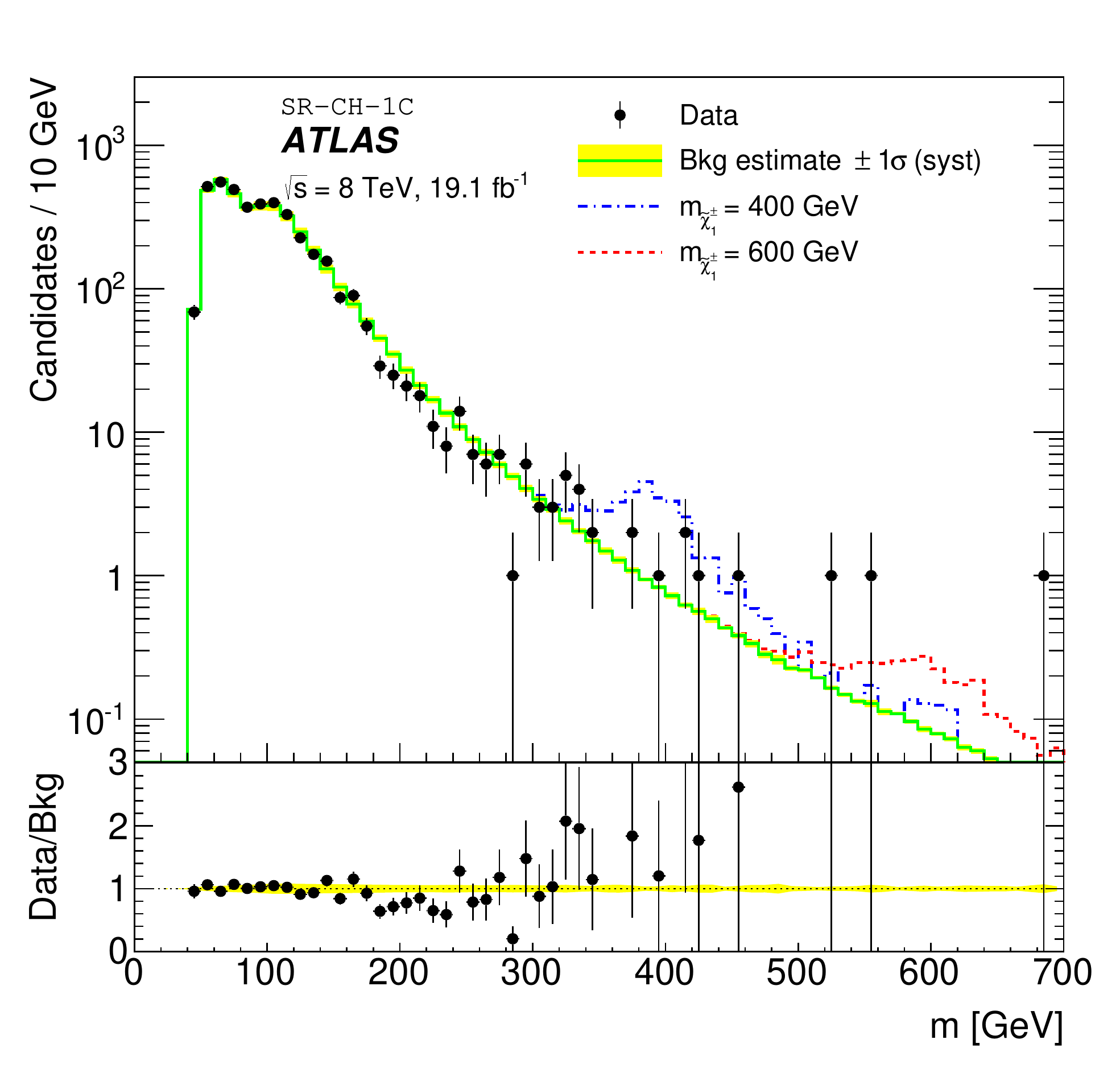}
 \caption{Reconstructed mass $m_\beta$ in observed data, background estimate and expected signal ($\tilde{\chi}^{\pm}_1$ masses of 400 and 600~\GeV) in the chargino search for the lower of the two masses ($m$) in the two-candidate signal region (top-left), for the one-loose-candidate signal region (top-right) and the one-candidate signal region (bottom).} 
  \label{fig:chargino_results}
\end{figure*}

\begin{figure*}[tb] 
  \centering 
  \includegraphics[width=.48\linewidth]{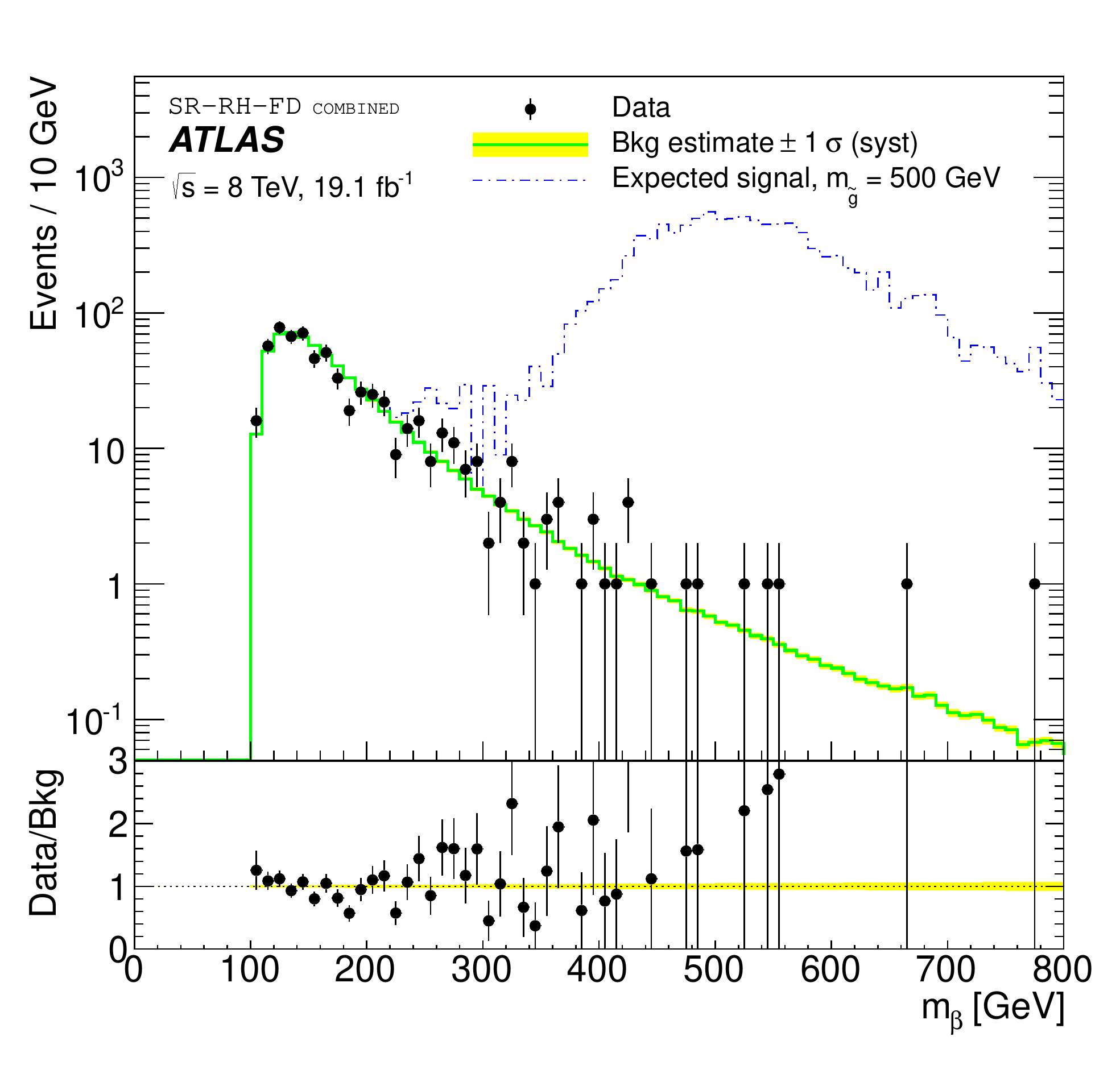} 
  \includegraphics[width=.48\linewidth]{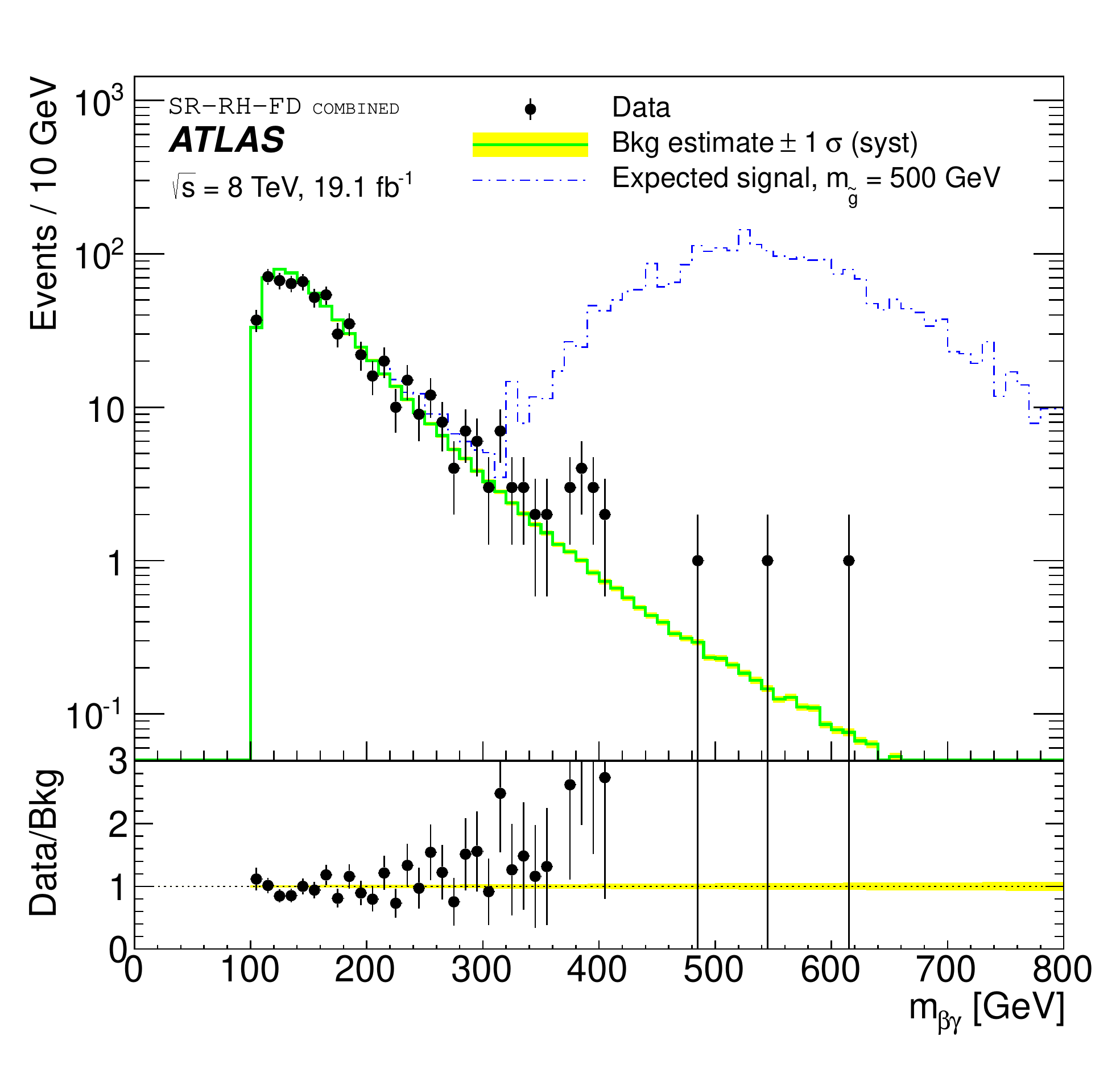} 
  \includegraphics[width=.48\linewidth]{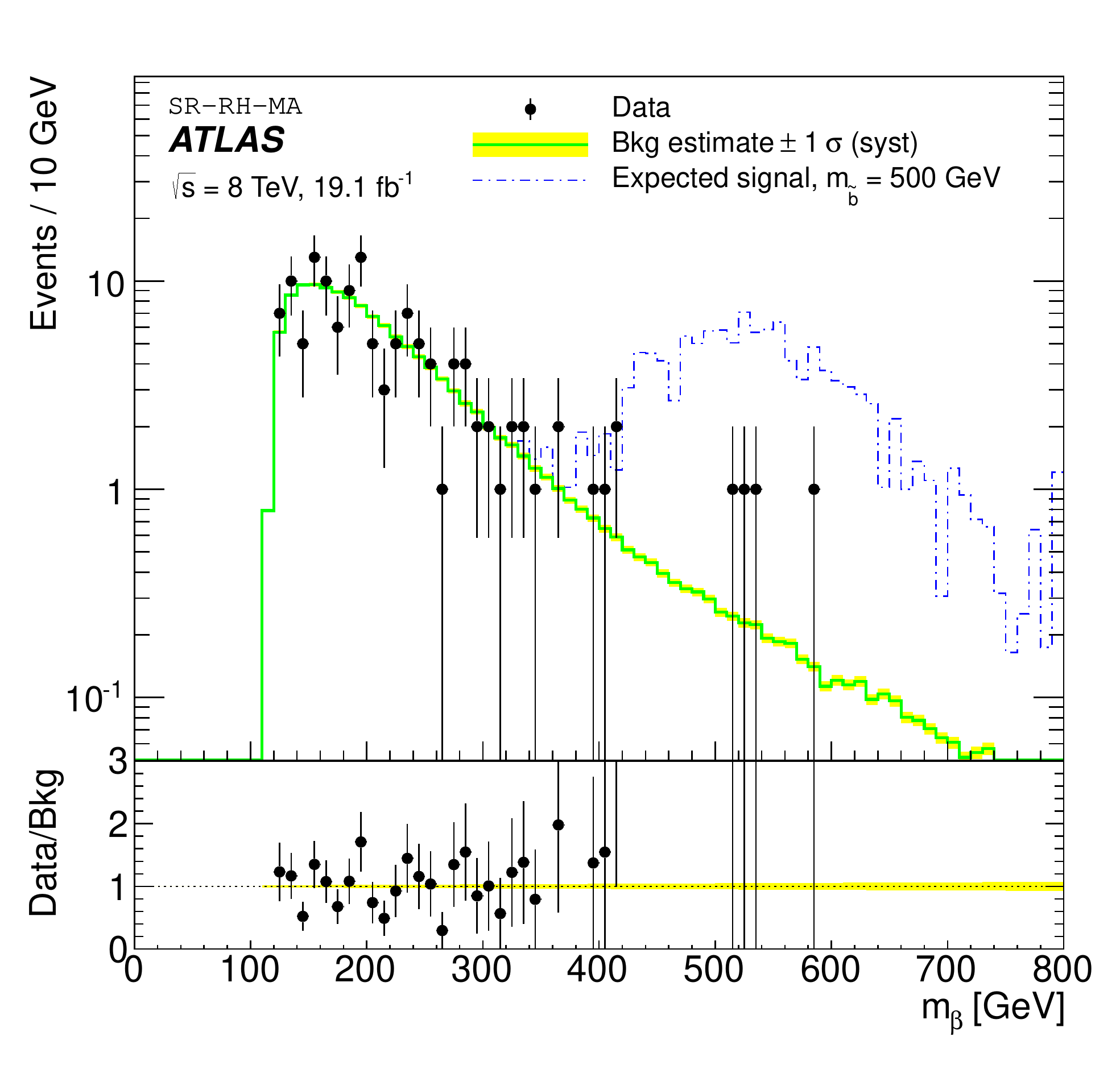}
  \includegraphics[width=.48\linewidth]{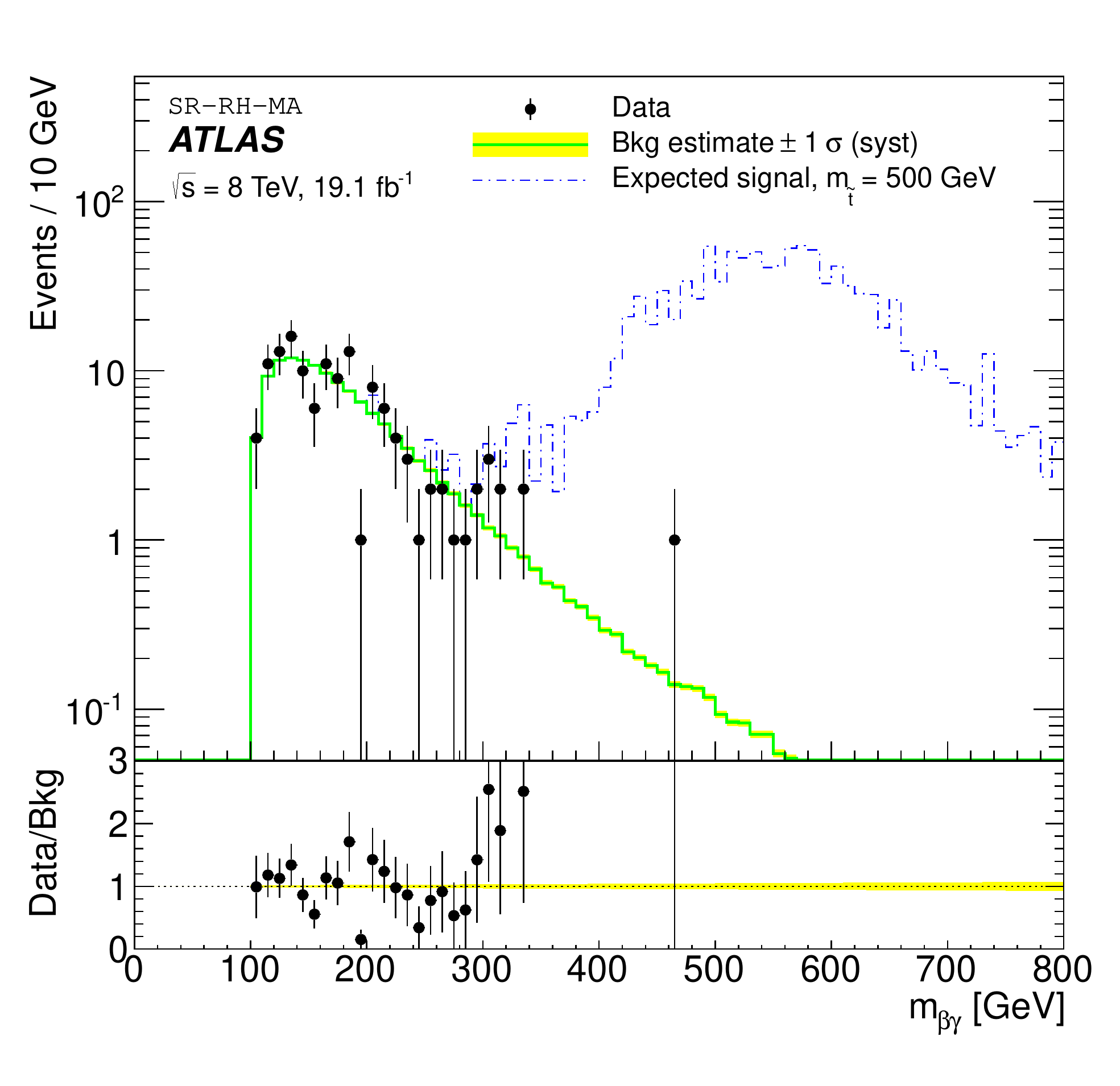} 
  \caption{Data and background estimates for the reconstructed mass based on time-of-flight, $m_{\beta}$, (top-left) and based on specific energy loss, $m_{\beta\gamma}$, (top-right) for \combined candidates in the full-detector 500~\GeV gluino \rhad search (\srrhfd), as well as $m_{\beta}$ (bottom-left) for the MS-agnostic 500~\GeV sbottom and $m_{\beta\gamma}$ (bottom-right) for the MS-agnostic 500~\GeV stop \rhad search (\srrhma). \label{fig:rhadronBkgDists}} 
\end{figure*}

The mass distributions observed in data together with the background estimate, its systematic uncertainty and examples of expected signal are shown in figures~\ref{fig:DSB}--\ref{fig:rhadronBkgDists} for the slepton, chargino and \rhad searches, respectively.

No indication of signal above the expected background is observed, and limits on new physics scenarios are set using the \CLs prescription~\cite{Read:2000ru}.

For each search, limits on the cross-section are calculated from the likelihood to observe the number of events found in each signal region with candidate mass $m_\beta$ (and $m_{\beta\gamma}$ for \rhads) above the required mass value, given the background estimate and signal efficiency. For each signal region the likelihood function is built assuming a Poisson counting model for the observed number of events with Gaussian constraints for the systematic uncertainties. For the slepton and the chargino searches a global extended likelihood, given by the product of the likelihood functions of the various signal regions, is used. The \CLs calculation uses a profile likelihood test statistic~\cite{Roostat}.

Mass limits are derived by comparing the obtained cross-section limits to the lower edge of the $\pm 1\sigma$ band around the theoretically predicted cross-section for each process. Examples of the observed and expected event yields, as well as efficiencies and uncertainties for data and some MC simulation signal samples, are shown in tables~\ref{tab:GMSBResults}--\ref{tab:RHADResults} for the various searches and signal regions.

\begin{table*}[tb]
   \footnotesize
   \begin{center}
    {\setlength{\tabcolsep}{0em}\begin{tabular*}{\linewidth}{@{\extracolsep{\fill}}clrrr}
      \hline
      \hline
      & $\stau_1$ mass [GeV]                  &           345 &           407 &           469 \\
      \hline
      \multirow{5}{*}{\begin{turn}{90}\srslcc\end{turn}}
      & Minimum $m_{\beta}$ requirement [GeV] &           240 &           270 &           320 \\
      & Expected signal                       &          12.5 &           5.1 &           2.1 \\
      & Efficiency                            & 0.28$\pm$0.01 & 0.29$\pm$0.01 & 0.28$\pm$0.01 \\
      & Estimated background                  & 0.43$\pm$0.05 & 0.25$\pm$0.03 & 0.10$\pm$0.01 \\
      & Observed                              &             0 &             0 &             0 \\
      \hline
      \multirow{5}{*}{\begin{turn}{90}\srslc\end{turn}}
      & Minimum $m_{\beta}$ requirement [GeV] &           240 &           280 &           320 \\
      & Expected signal                       &           8.5 &           3.5 &           1.5 \\
      & Efficiency                            & 0.19$\pm$0.01 & 0.20$\pm$0.01 & 0.21$\pm$0.01 \\
      & Estimated background                  &      49$\pm$5 &      27$\pm$3 &      15$\pm$1 \\
      & Observed                              &            47 &            28 &            20 \\
      \hline
      & Cross-section limit [fb]              &          0.52 &          0.50 &          0.54 \\
      \hline
      \hline 
    \end{tabular*}}
    \caption{Observed and expected event yields, as well as efficiencies and uncertainties for three MC simulation signal samples, in the two signal regions used in the GMSB slepton search. Cross-section upper limits are stated at 95\% CL.} \label{tab:GMSBResults}
  \end{center}
\end{table*}

\begin{table*}[tb]
   \footnotesize
   \begin{center}
    {\setlength{\tabcolsep}{0em}\begin{tabular*}{\linewidth}{@{\extracolsep{\fill}}clrrr}
      \hline
      \hline
        & $\tilde{q}$ mass [GeV]                 &            1200 &            2000 &            3000 \\
        & $\tilde{g}$ mass [GeV]                 &            1600 &            1400 &            1600 \\
    \hline
    \multirow{5}{*}{\begin{turn}{90}\srslcc\end{turn}}
        & Minimum $m_{\beta}$ requirement [\GeV] &             190 &             190 &             190 \\
        & Expected signal                        &            96.5 &             6.9 &             0.5 \\
        & Efficiency                             &   0.27$\pm$0.01 &   0.30$\pm$0.01 &   0.19$\pm$0.01 \\
        & Estimated background                   &   1.36$\pm$0.14 &   1.36$\pm$0.14 &   1.36$\pm$0.14 \\
        & Observed                               &               0 &               0 &               0 \\
    \hline
    \multirow{5}{*}{\begin{turn}{90}\srslc\end{turn}}
        & Minimum $m_{\beta}$ requirement [\GeV] &             210 &             210 &             210 \\
        & Expected signal                        &            66.9 &             4.9 &             0.5 \\
        & Efficiency                             & 0.190$\pm$0.007 & 0.207$\pm$0.007 & 0.181$\pm$0.006 \\
        & Estimated background                   &        80$\pm$7 &        80$\pm$7 &        80$\pm$7 \\
        & Observed                               &              73 &              73 &              73 \\
    \hline
        & Cross-section limit [fb]               &            0.55 &            0.49 &            0.78 \\
      \hline
      \hline 
    \end{tabular*}}
    \caption{Observed and expected event yields, as well as efficiencies and uncertainties for three MC simulation signal samples, in the three signal regions used in the \lsusy slepton search. Cross-section upper limits are stated at 95\% CL.} \label{tab:LSResults}
  \end{center}
\end{table*}

\begin{table*}[tb]
   \footnotesize
   \begin{center}
    {\setlength{\tabcolsep}{0em}\begin{tabular*}{\linewidth}{@{\extracolsep{\fill}}clrrr}
      \hline
      \hline
      & $\tilde{\chi}^{\pm}_{1}$ mass [GeV]   &             500 &             600 &             700 \\
      \hline
      \multirow{5}{*}{\begin{turn}{90}\srchcc\end{turn}}
      & Minimum $m_{\beta}$ requirement [GeV] &             350 &             420 &             480 \\
      & Expected signal                       &            16.9 &             4.9 &             1.5 \\
      & Efficiency                            & 0.061$\pm$0.003 & 0.054$\pm$0.002 & 0.047$\pm$0.002 \\
      & Estimated background                  & 0.053$\pm$0.006 & 0.018$\pm$0.003 & 0.008$\pm$0.001 \\
      & Observed                              &               0 &               0 &               0 \\
      \hline
      \multirow{5}{*}{\begin{turn}{90}\srchcl\end{turn}}
      & Minimum $m_{\beta}$ requirement [GeV] &             300 &             330 &             420 \\
      & Expected signal                       &            35.0 &            10.7 &             3.3 \\
      & Efficiency                            & 0.126$\pm$0.006 & 0.118$\pm$0.005 & 0.109$\pm$0.005 \\
      & Estimated background                  &    29.6$\pm$0.3 &    21.1$\pm$0.3 &     8.6$\pm$0.3 \\
      & Observed                              &              37 &              31 &              12 \\
      \hline
      \multirow{5}{*}{\begin{turn}{90}\srchc\end{turn}}
      & Minimum $m_{\beta}$ requirement [GeV] &             340 &             430 &             450 \\
      & Expected signal                       &            9.21 &            2.95 &            0.99 \\
      & Efficiency                            & 0.033$\pm$0.002 & 0.033$\pm$0.002 & 0.032$\pm$0.002 \\
      & Estimated background                  &  14.14$\pm$0.67 &   4.85$\pm$0.21 &   3.91$\pm$0.16 \\
      & Observed                              &              14 &               6 &               6 \\
      \hline
      & Cross-section limit [fb]              &            2.18 &            3.31 &            2.62 \\
      \hline
      \hline 
    \end{tabular*}}
    \caption{Observed and expected event yields, as well as efficiencies and uncertainties for three MC simulation signal samples, in the three signal regions used in the chargino search. Cross-section upper limits are stated at 95\% CL.} \label{tab:CHARResults}
  \end{center}
\end{table*}

\begin{table*}[tb]
   \footnotesize
   \begin{center}
    {\setlength{\tabcolsep}{0em}\begin{tabular*}{\linewidth}{@{\extracolsep{\fill}}clrrr}
      \hline
      \hline
      & \rhad type / mass [GeV]                     & $\tilde{g}$ / 1300 & $\tilde{b}$ / 800 & $\tilde{t}$ / 900 \\
      \hline
      \multirow{6}{*}{\begin{turn}{90}\srrhma\end{turn}}
      & Minimum $m_{\beta\gamma}$ requirement [GeV] &              785.1 &             560.3 &             612.9 \\
      & Minimum $m_{\beta}$ requirement [GeV]       &              746.9 &             512.7 &             565.5 \\
      & Expected signal                             &               3.09 &              4.63 &              2.54 \\
      & Efficiency                                  &      0.10$\pm$0.01 &   0.084$\pm$0.009 &     0.12$\pm$0.01 \\
      & Estimated background                        &    0.010$\pm$0.001 &     0.27$\pm$0.02 &     0.18$\pm$0.02 \\
      & Observed                                    &                  0 &                 0 &                 0 \\
      \hline
      & Cross-section limit [fb]                    &               1.53 &               1.8 &              1.23 \\
      \hline
      \hline
      &                                             &                    &                   &                   \\
      \hline
      \hline
      & \rhad type / mass [GeV]                     & $\tilde{g}$ / 1300 & $\tilde{b}$ / 800 & $\tilde{t}$ / 900 \\
      \hline
      \multirow{6}{*}{\begin{turn}{90}\srrhfd\end{turn}}
      & Minimum $m_{\beta\gamma}$ requirement [GeV] &              785.1 &             560.3 &             612.9 \\
      & Minimum $m_{\beta}$ requirement [GeV]       &              746.9 &             512.7 &             565.5 \\
      & Expected signal                             &               3.50 &              5.90 &              3.50 \\
      & Efficiency                                  &      0.11$\pm$0.01 &     0.11$\pm$0.01 &     0.17$\pm$0.02 \\
      & Estimated background                        &    0.051$\pm$0.006 &     0.73$\pm$0.06 &     0.40$\pm$0.03 \\
      & Observed                                    &                  0 &                 1 &                 0 \\
      \hline
      & Cross-section limit [fb]                    &               1.33 &              1.80 &              0.84 \\
      \hline
      \hline 
    \end{tabular*}} 
    \caption{Observed and expected event yields, as well as efficiencies and uncertainties for three MC simulation signal samples, in the \rhad searches. Cross-section upper limits are stated at 95\% CL.} \label{tab:RHADResults}
  \end{center}
\end{table*}

\subsection{Slepton limits}

\begin{figure*}[tb]
  \centering
  \includegraphics[width=0.48\linewidth]{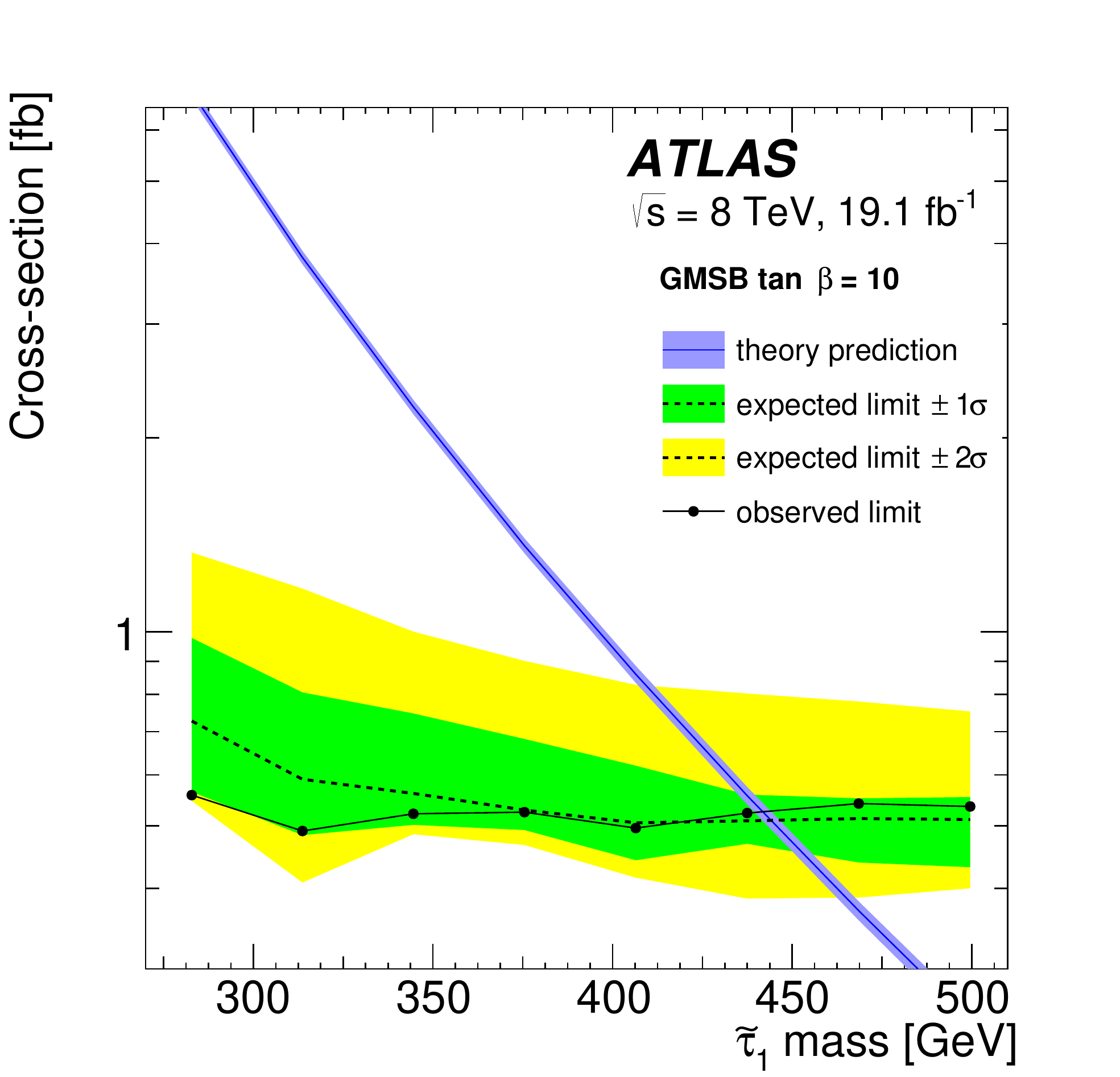}
  \includegraphics[width=0.48\linewidth]{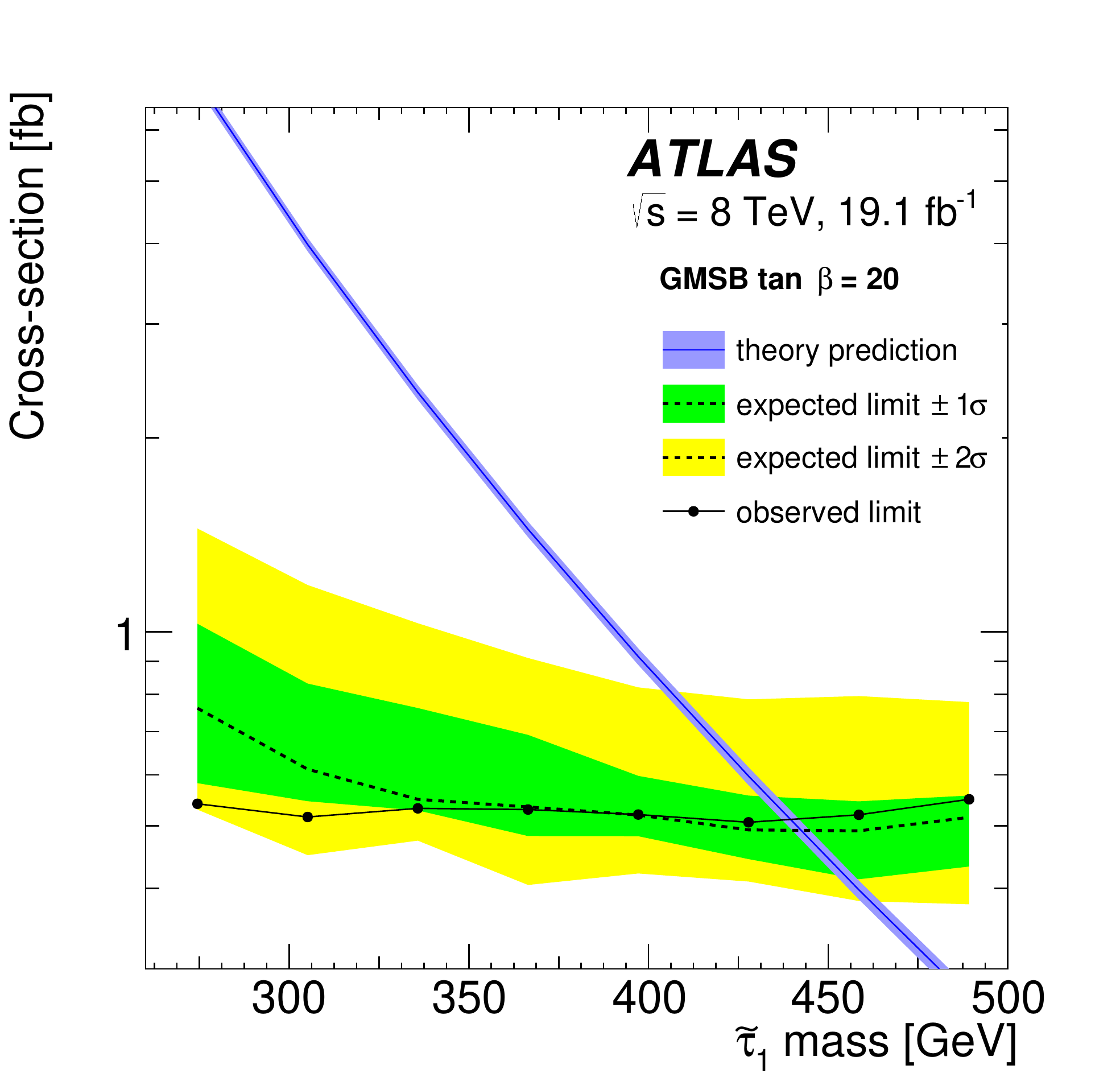}
  \includegraphics[width=0.48\linewidth]{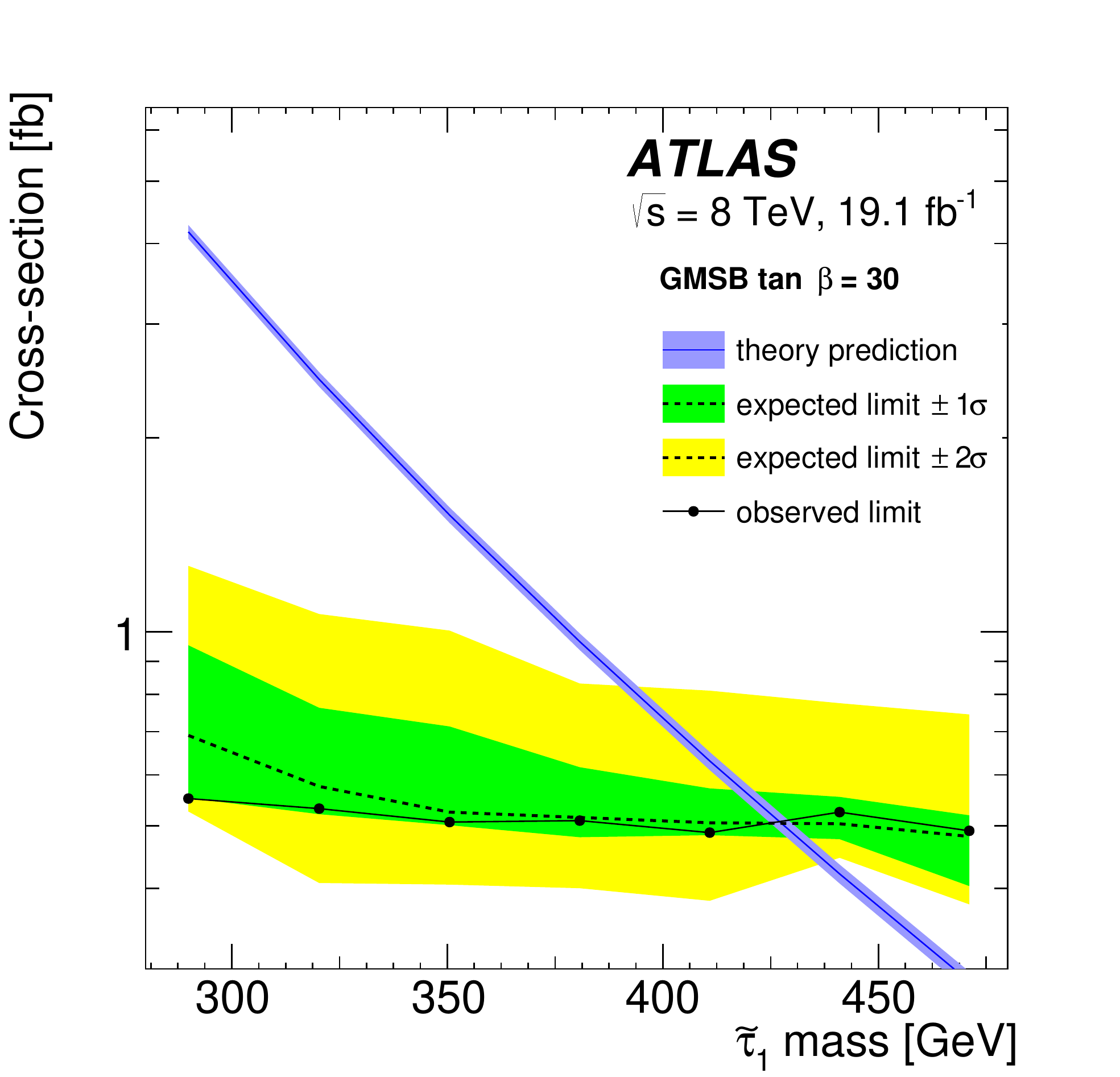}
  \includegraphics[width=0.48\linewidth]{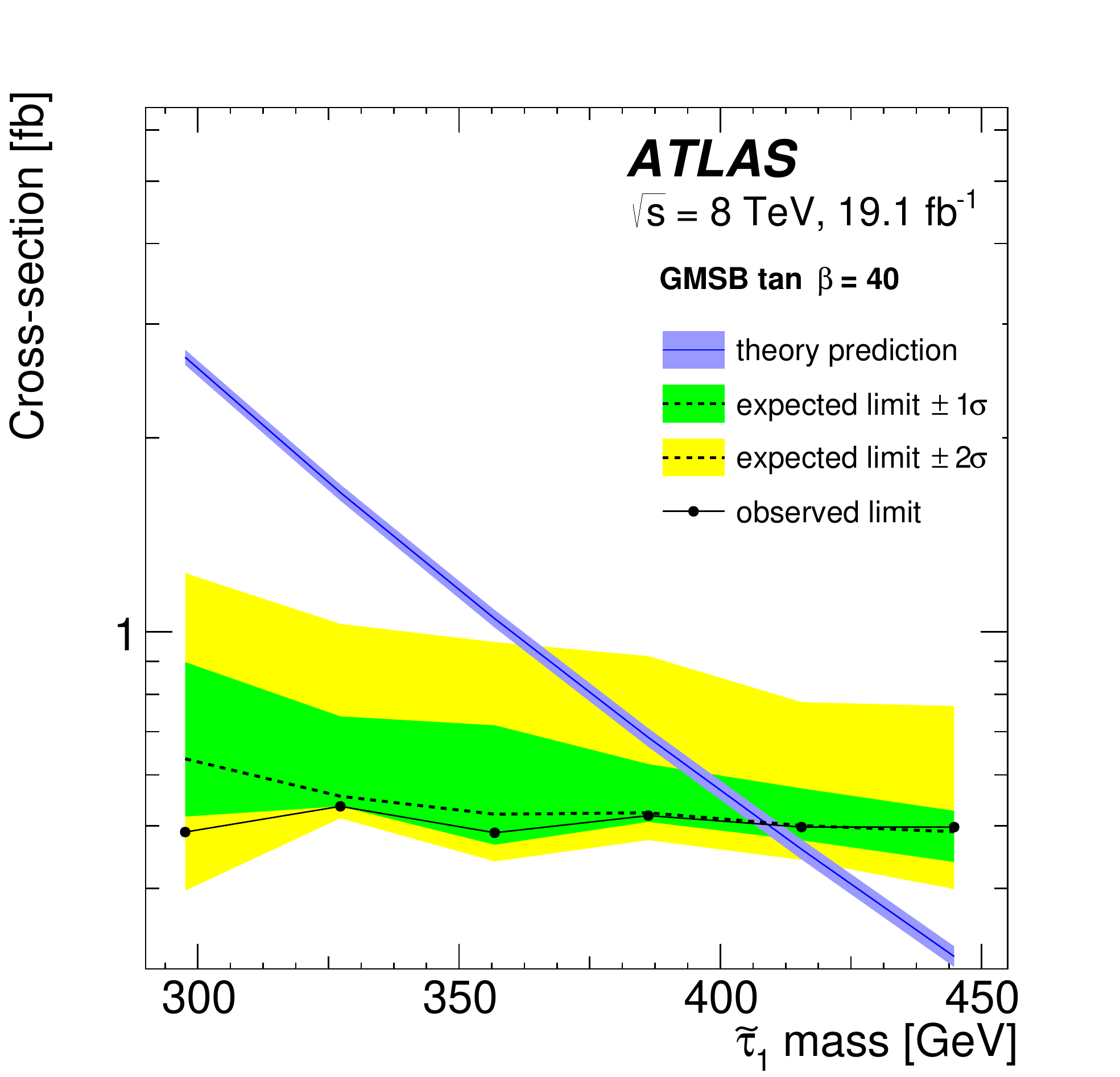}
  \includegraphics[width=0.48\linewidth]{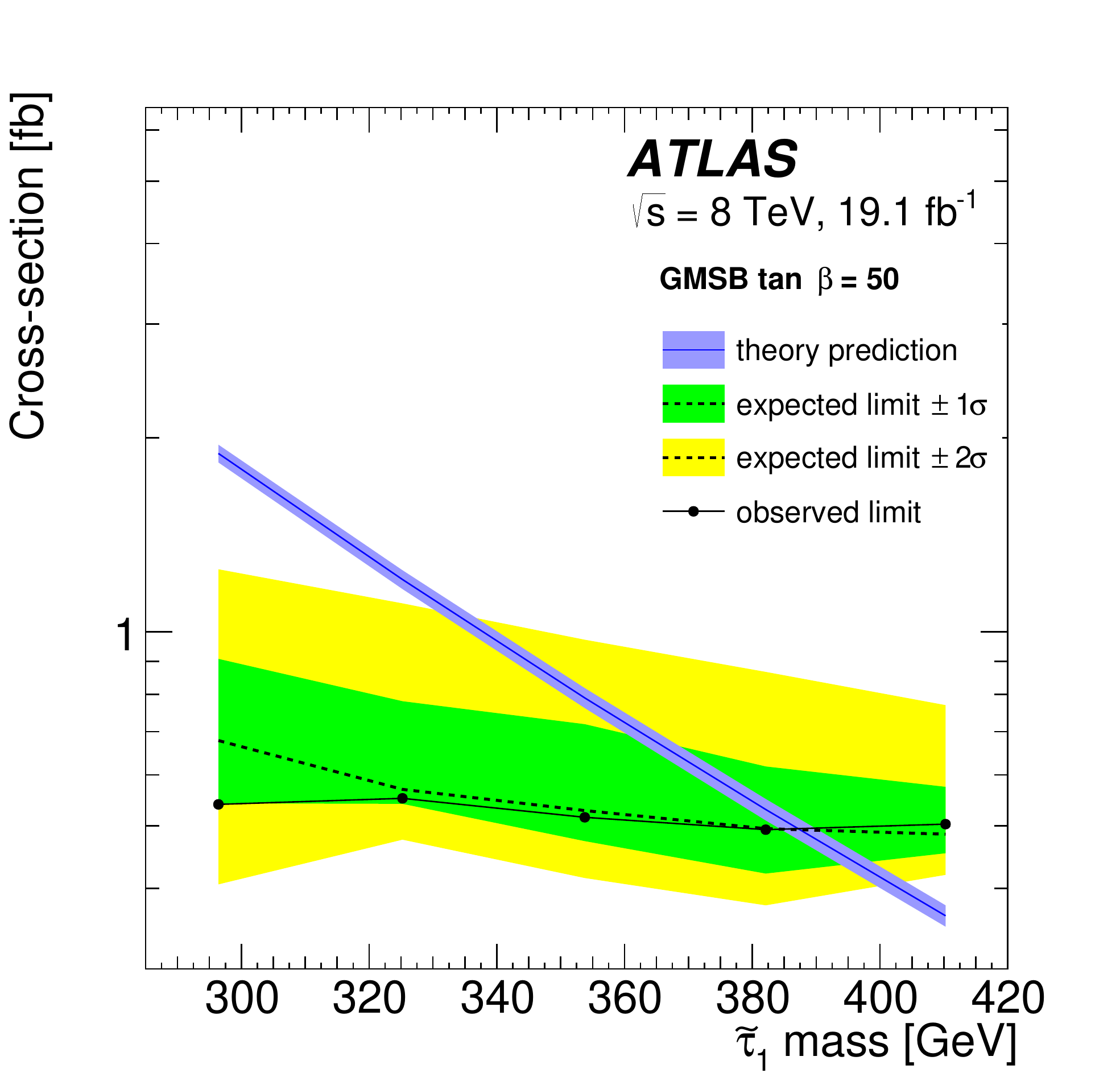}
  \caption{Cross-section upper limits as a function of the mass of the lightest stau for the GMSB models organised by $\tan\beta$: 10 and 20 first row, 30 and 40 second row and 50 last row. The expected limit is drawn as a dashed black line with $\pm 1\sigma$ and $\pm 2\sigma$ uncertainty bands drawn in green and yellow, respectively. The observed limit is shown as solid black line with markers. The theoretical cross-section prediction is shown as a solid blue line with a shaded $\pm 1\sigma$ uncertainty band.} \label{fig:limit_by_tanb}
\end{figure*}

The resulting production cross-section limits at 95\% confidence level (CL) in the GMSB scenario as a function of the $\stau_1$ mass are presented in figure~\ref{fig:limit_by_tanb} and compared to theoretical predictions. 

A long-lived $\stau_1$ in GMSB models with $N_5=3$, $m_{\rm{messenger}}=250~\TeV$ and $\mathrm{sign}(\mu)=1$ is excluded at 95\% CL up to masses of \gmsbAllExclusion~\GeV for $\tan\beta$ = 10, 20, 30, 40 and 50, respectively. 

\begin{figure*}[tb]
  \centering
  \includegraphics[width=0.48\linewidth]{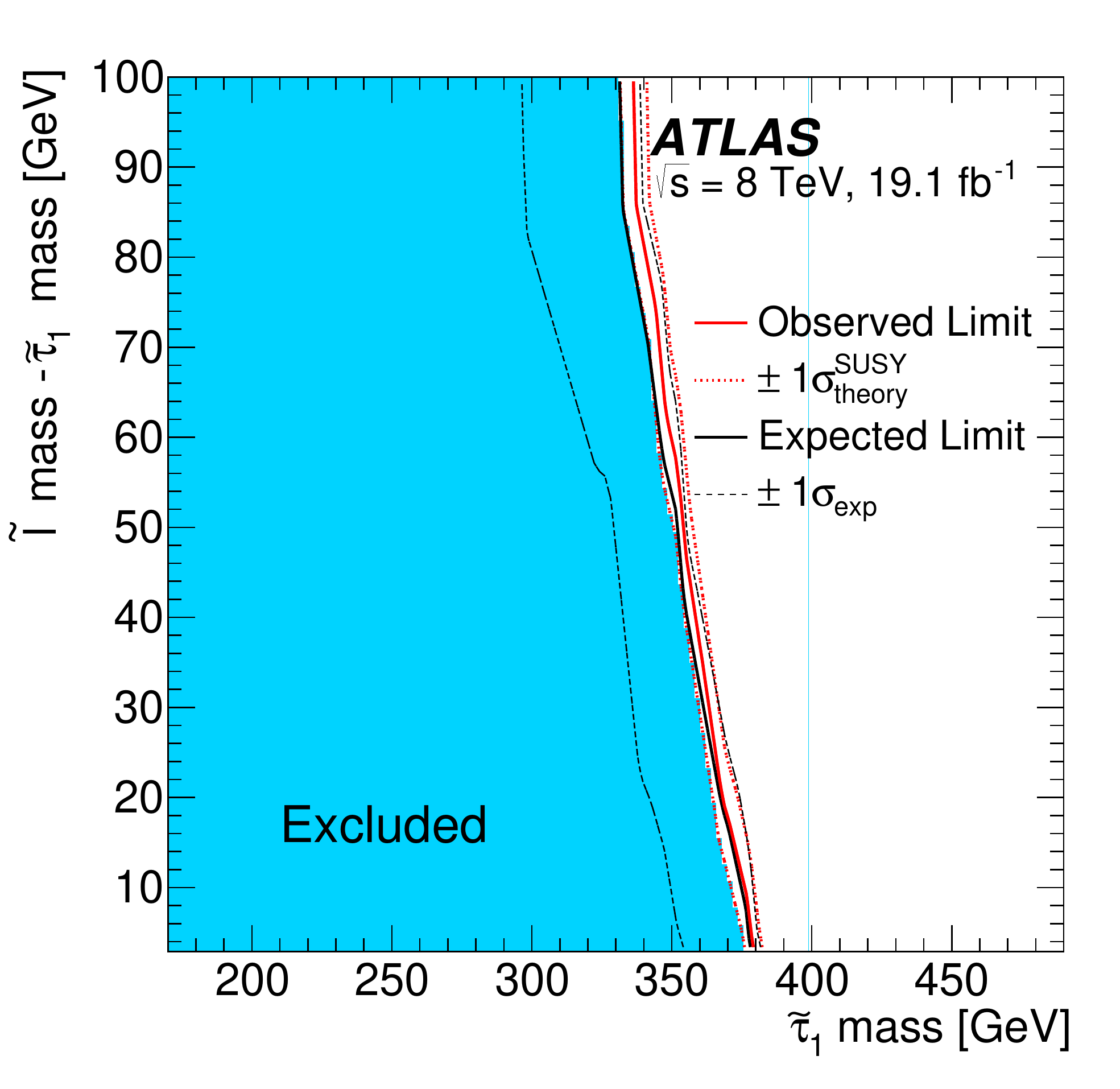}
  \caption{95\% CL excluded regions for directly produced sleptons in the plane $m(\tilde{\ell})$--$m(\stau_1)$ vs. $m(\stau_1)$. The excluded region is is shown in blue. The expected limit is drawn as a solid black line with a $\pm 1 \sigma$ uncertainty band drawn as dashed black lines. The observed limit is shown as solid red line with a $\pm 1 \sigma$ uncertainty band drawn as dashed red lines.}
  \label{fig:limits2d-DY}
\end{figure*}

\begin{figure*}[tb]
  \centering
  \includegraphics[width=0.48\linewidth]{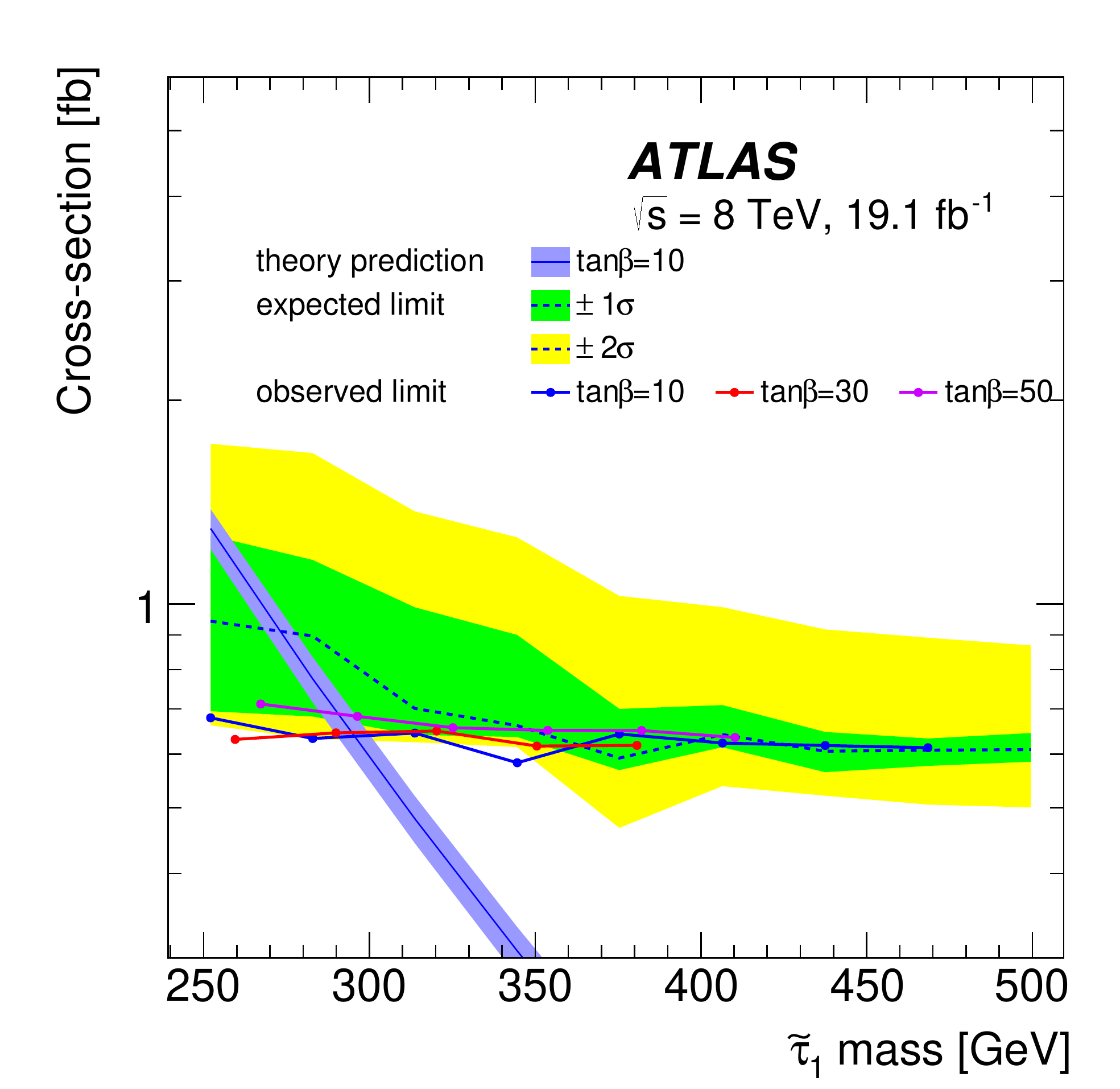}
  \caption{Cross-section upper limits as a function of the $\stau_1$ mass for direct $\stau_1$ production and three values of $\tan\beta$. The expected limit for $\tan\beta=10$ is drawn as a dashed black line with $\pm 1\sigma$ and $\pm 2\sigma$ uncertainty bands drawn in green and yellow, respectively. The observed limits for three values of $\tan\beta$ are shown as solid lines with markers. The theoretical cross-section prediction for $\tan\beta=10$ is shown as a coloured $\pm 1\sigma$ band, and does not vary significantly for the other $\tan\beta$ values.
}
  \label{fig:limitsSTAU-STAU}
\end{figure*}

Limits on the rates of specific production mechanisms are obtained by repeating the analysis on subsets of the GMSB samples corresponding to each production mode. For GMSB models with parameters in this range, strong production of squarks and gluinos is suppressed due to their large masses. Directly produced sleptons constitute 30--63\% of the GMSB cross-section, and the corresponding $\stau_1$ production rates depend only on the $\stau_1$ mass and the mass difference between the right-handed $\tilde{e}$ (or $\tilde{\mu}$) and the $\stau_1$. Thus the same analysis constrains a simple model with only pair-produced sleptons which are long-lived, or which themselves decay to long-lived sleptons of another flavour. Such direct production is excluded at 95\% CL up to $\stau_1$ masses of 373 to 330~\GeV for models with slepton mass splittings of 2.7--93~\GeV. The slepton direct-production limits are shown in figure~\ref{fig:limits2d-DY}. Figure~\ref{fig:limitsSTAU-STAU} shows the cross-section limits on direct $\stau_1$ production for the case where the mass splitting from the other sleptons is very large. As the theoretical prediction as well as the according uncertainty bands overlap almost entirely for various values of $\tan\beta$, only the curve for $\tan\beta = 10$ is shown. Masses below 286~\GeV are excluded if only $\stau_1$ is produced. The values for direct $\stau_1$-only production are also used in figure~\ref{fig:limits2d-DY} at very high mass splittings.

\begin{figure*}[tb]
  \centering
  \includegraphics[width=0.48\linewidth]{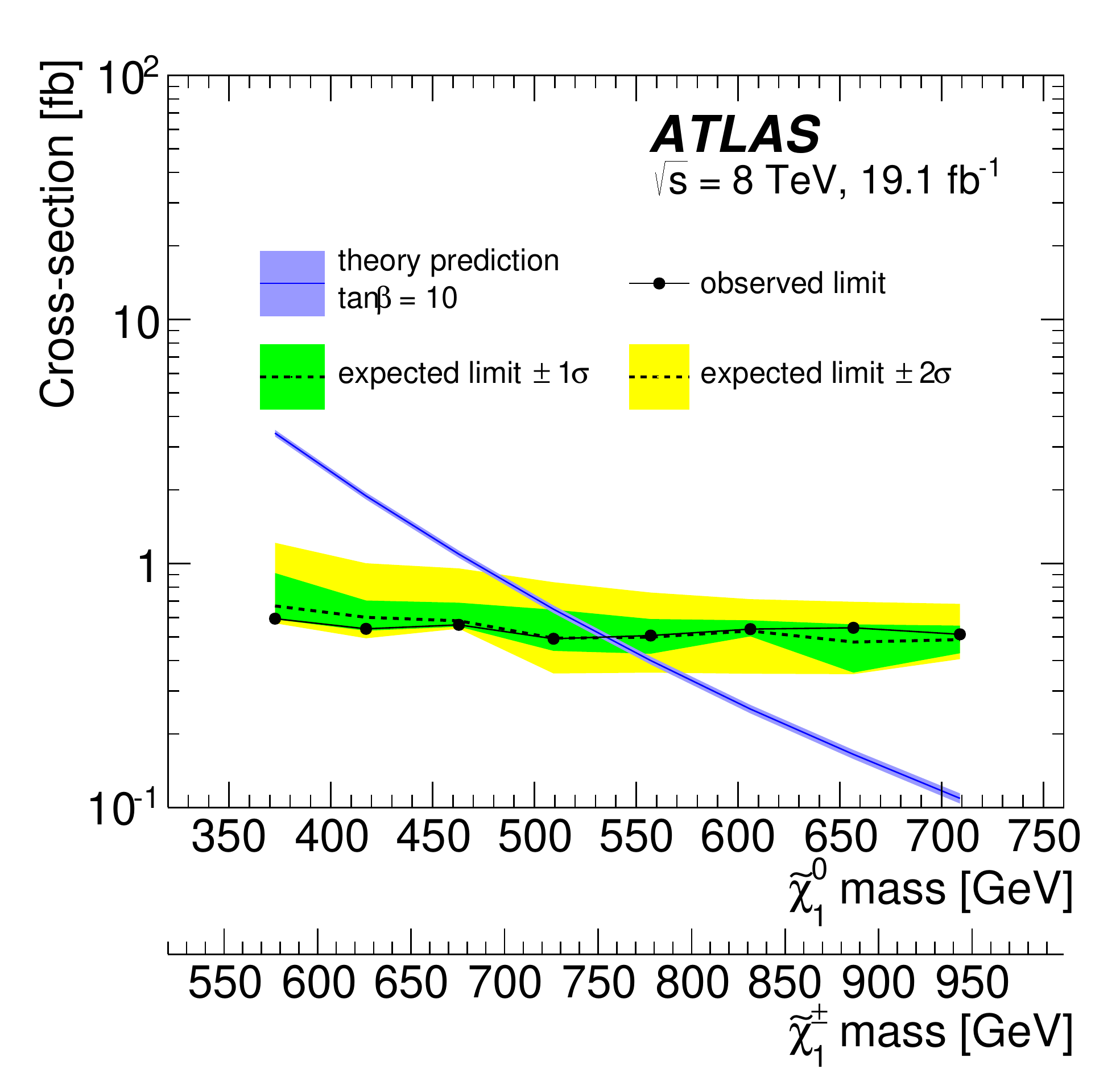}
  \caption{Cross-section upper limits as a function of the $\tilde{\chi}_1$ mass for $\stau_1$ sleptons resulting from the decay of directly produced charginos and neutralinos in GMSB. Observed limits are given as a solid black line with markers. Expected limits for $\tan\beta=10$ are drawn as a dashed black line with $\pm 1\sigma$ and $\pm 2\sigma$ uncertainty bands drawn in green and yellow, respectively. The theoretical cross-section prediction (dominated by $\tilde{\chi}^0_1 \tilde{\chi}^+_1$ production) is shown as a coloured $\pm 1\sigma$ band. Depending on the hypothesis and to a small extent on $\tan\beta$, in these models, the chargino mass is 210 to 260~\GeV higher than the neutralino mass.}
  \label{fig:limitsCHI-EW}
\end{figure*}

Finally, in the context of the GMSB model, 30--50\% of the GMSB cross-section arises from direct production of charginos and neutralinos (dominated by $\tilde{\chi}^0_1 \tilde{\chi}^{\pm}_1$ production) and subsequent decay to $\stau_1$. Figure~\ref{fig:limitsCHI-EW} shows the 95\% CL lower limits on the $\tilde{\chi}^0_1$ and $\tilde{\chi}^\pm_1$ mass when the final decay product is a long-lived $\stau_1$. In the samples used to derive these limits, the $\tilde{\chi}^0_1$ and $\tilde{\chi}^{\pm}_1$ masses are closely related by GMSB, as represented by the values on the two $x$-axes. The mass of the $\stau_1$ decreases with increasing $\tan\beta$ and increases with the $\tilde{\chi}^0_1$ and $\tilde{\chi}^\pm_1$ masses. At low $\tilde{\chi}^0_1$ and $\tilde{\chi}^\pm_1$ masses and large $\tan\beta$, the cross-section limits are thus affected by the amount of background in the $\stau_1$ mass search region, which starts at 120~\GeV for $\tan\beta=50$ and at 170~\GeV for $\tan\beta=10$. The cross-section limits exclude $\tilde{\chi}^0_1$ masses below 537~\GeV, with corresponding $\tilde{\chi}^\pm_1$ masses 210--260~\GeV higher.

\begin{figure*}[tb]
   \centering
  \includegraphics[width=0.48\linewidth]{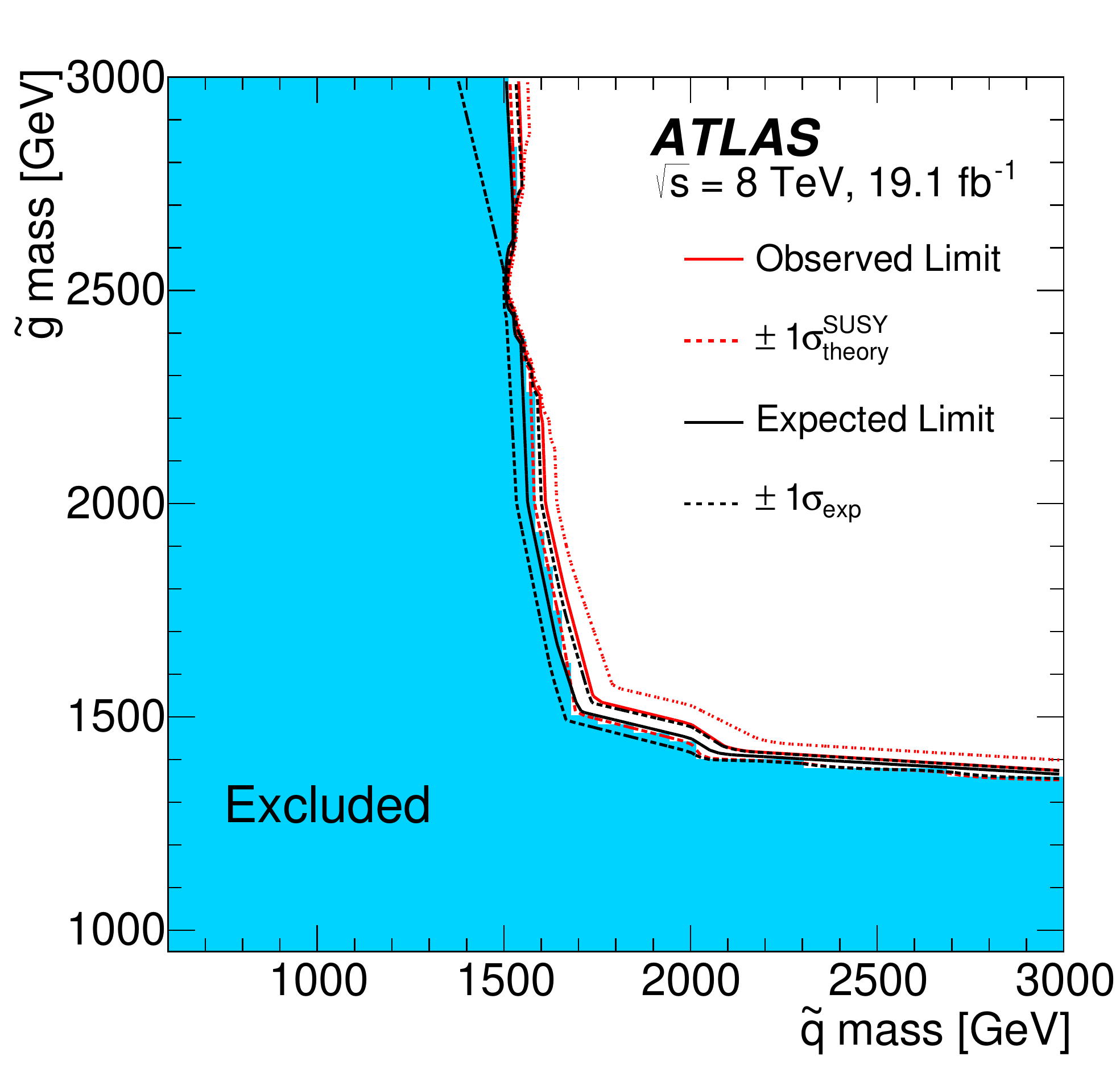}
  \caption{95\% CL excluded regions of squark mass and gluino mass in the \lsusy models. The excluded region is is shown in blue. The expected limit is drawn as a solid black line with a $\pm 1 \sigma$ uncertainty band drawn as dashed black lines. The observed limit is shown as solid red line with a $\pm 1 \sigma$ uncertainty band drawn as dashed red lines.}
  \label{fig:ls_limit}
\end{figure*}

Limits on \lsusy scenarios are set on squarks and gluinos decaying to long-lived sleptons within the \lsusy model. The exclusion region in the plane $m(\tilde{g})$ vs. $m(\tilde{q})$ is shown in figure~\ref{fig:ls_limit}. Squark and gluino masses are excluded at 95\% CL up to a mass of \limitleptosquark and \limitleptogluino~\GeV, respectively, in simplified \lsusy models where sleptons are stable and degenerate, with a mass of 300~\GeV, and all neutralinos (except $\tilde{\chi}^0_1$) and charginos are decoupled.

\subsection{Chargino limits}

\begin{figure*}[tb]
 \centering
 \includegraphics[width=0.48\linewidth]{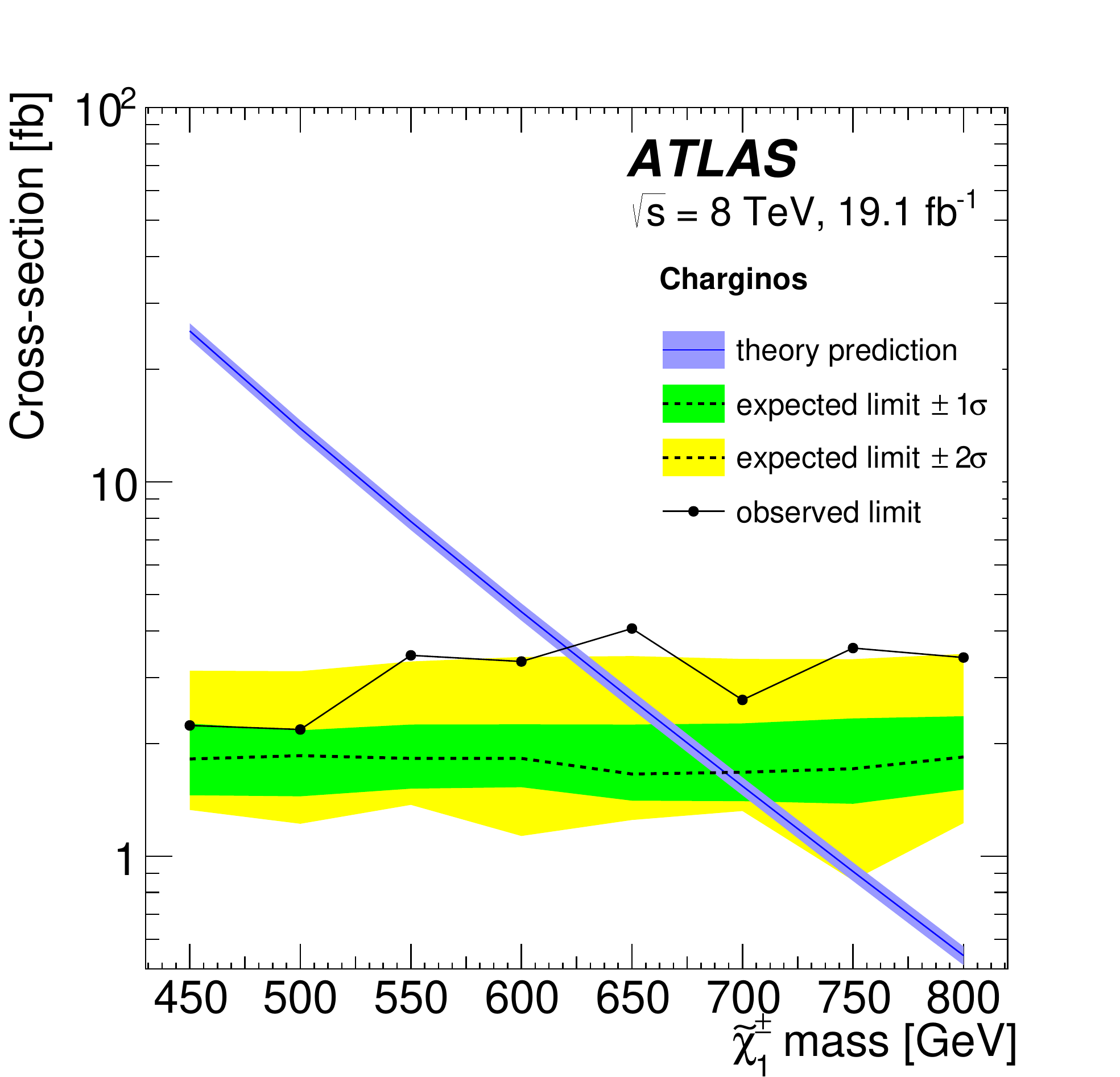}
\caption{Cross-section upper limits for various chargino masses in stable-chargino models. The expected limit is drawn as a dashed black line with $\pm 1\sigma$ and $\pm 2\sigma$ uncertainty bands drawn in green and yellow, respectively. The observed limit is shown as solid black line with markers. The theoretical cross-section prediction is shown as a solid blue line with a shaded $\pm 1\sigma$ uncertainty band.} 
 \label{fig:char_limit}
\end{figure*}

Limits are set on long-lived charginos, which are nearly degenerate with the lightest neutralino in simplified SUSY models. The production cross-section limits at 95\% CL in this scenario as a function of the $\tilde{\chi}^{\pm}_1$ mass are presented in figure~\ref{fig:char_limit} and compared to theoretical predictions. Masses below \limitcharginos~\GeV are excluded. The observed cross-section limit is found to be consistently one or two standard deviations ($\sigma$) above the expected limit, due to an excess of data events relative to the background estimate in \srchcl, as can be seen in figure~\ref{fig:chargino_results} (top-right).

\subsection{\rhad limits}

\begin{figure*}[tb]
 \centering
  \includegraphics[width=0.48\linewidth]{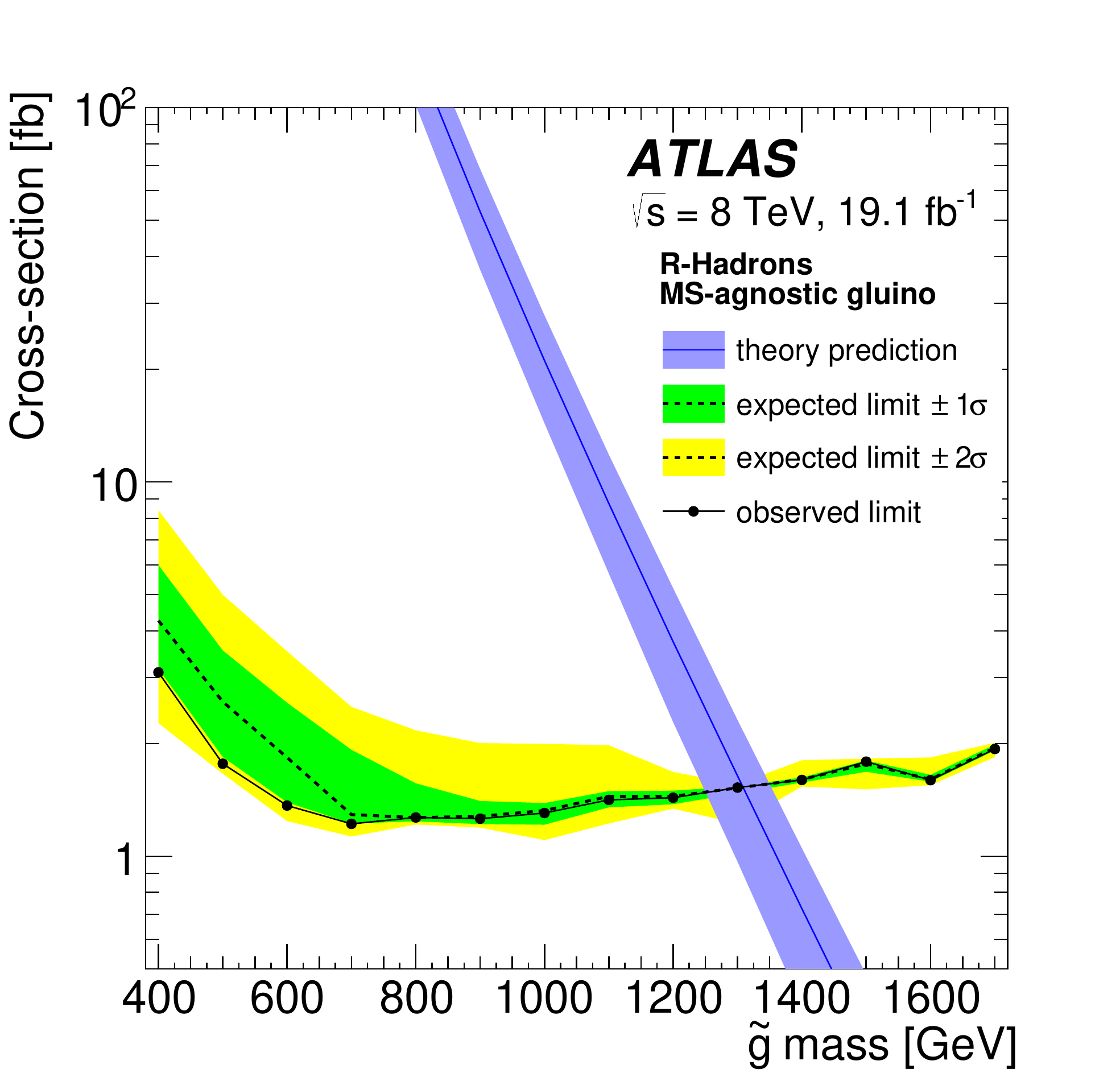}
  \includegraphics[width=0.48\linewidth]{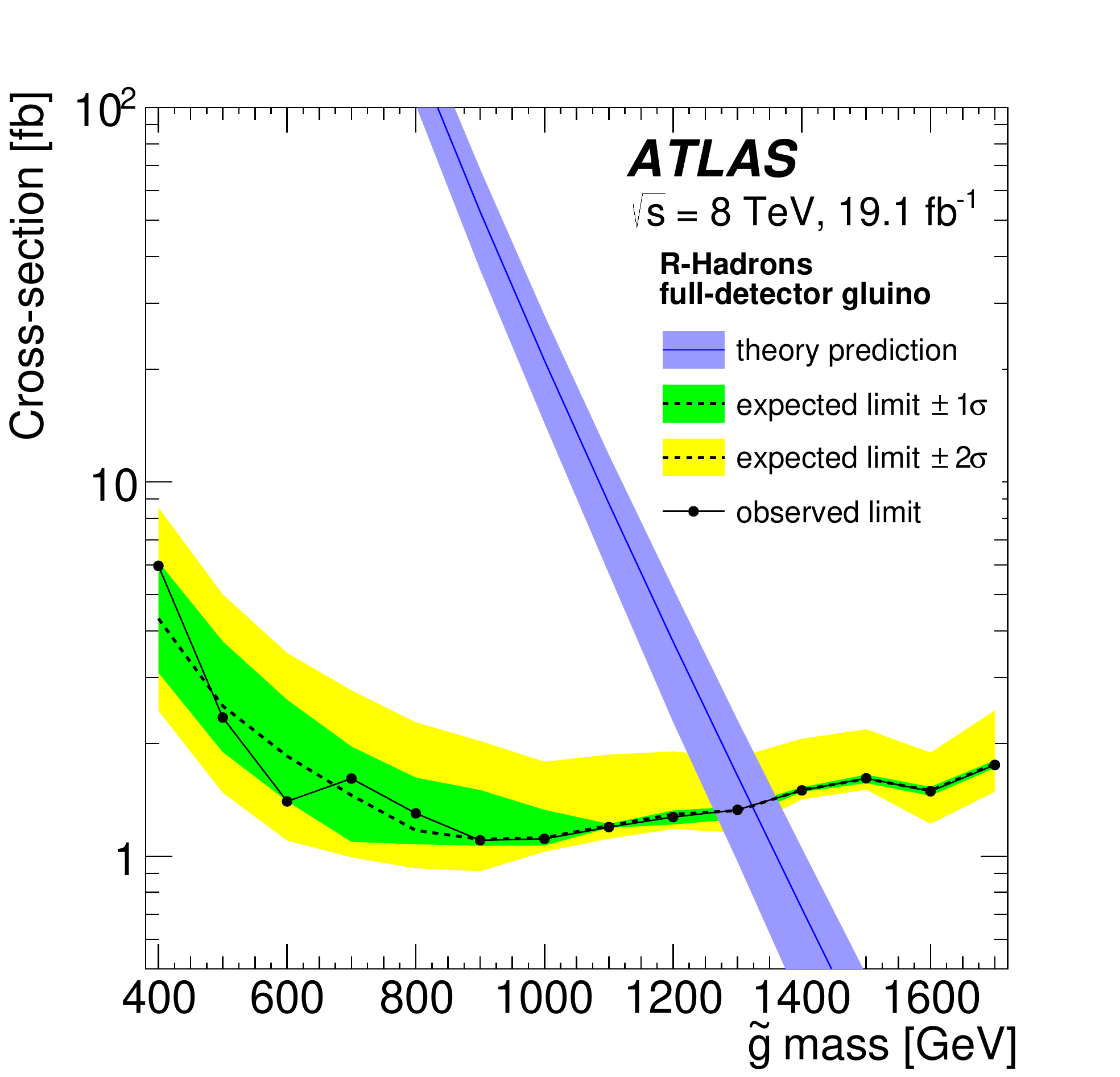}
  \includegraphics[width=0.48\linewidth]{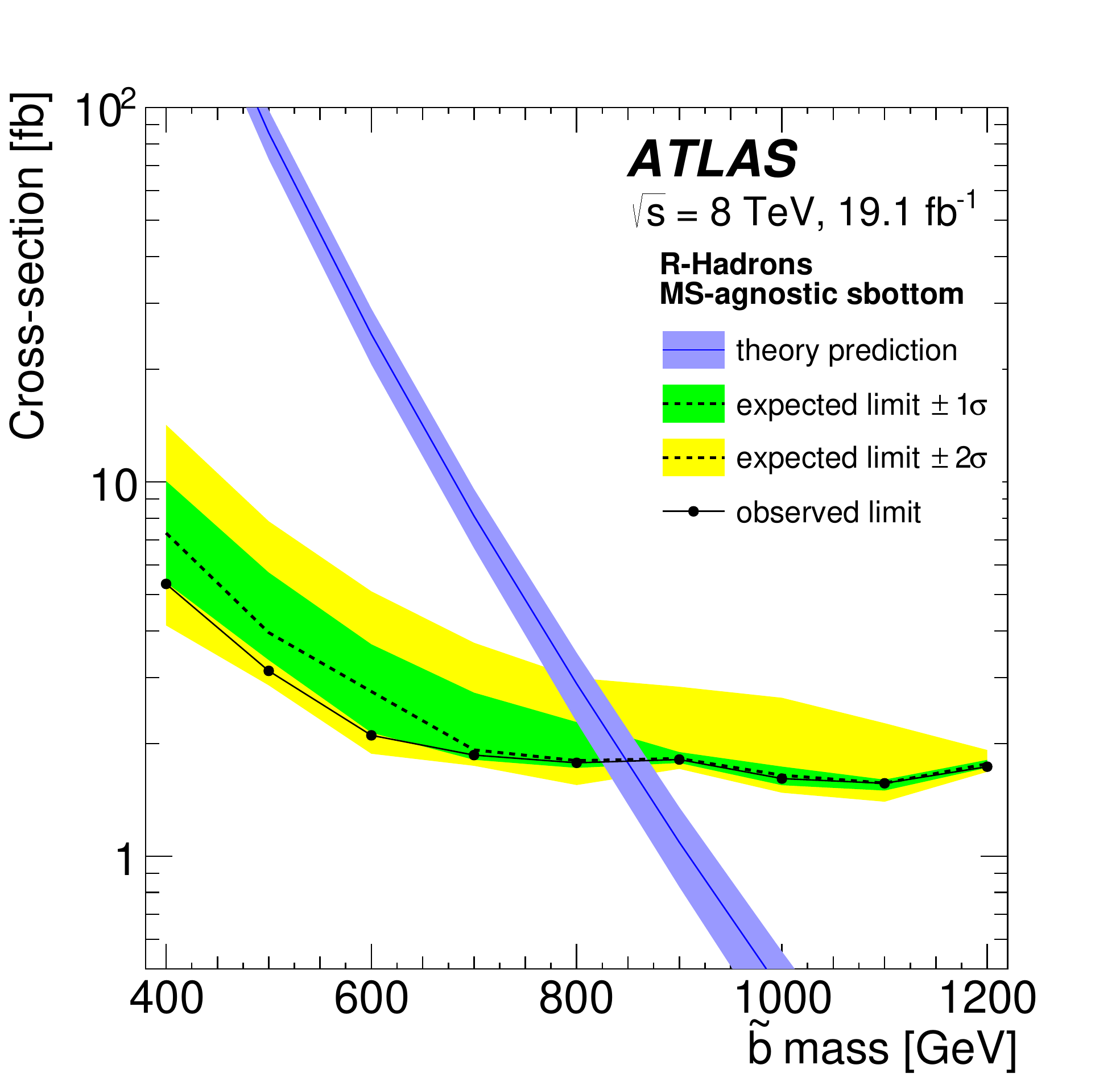}
  \includegraphics[width=0.48\linewidth]{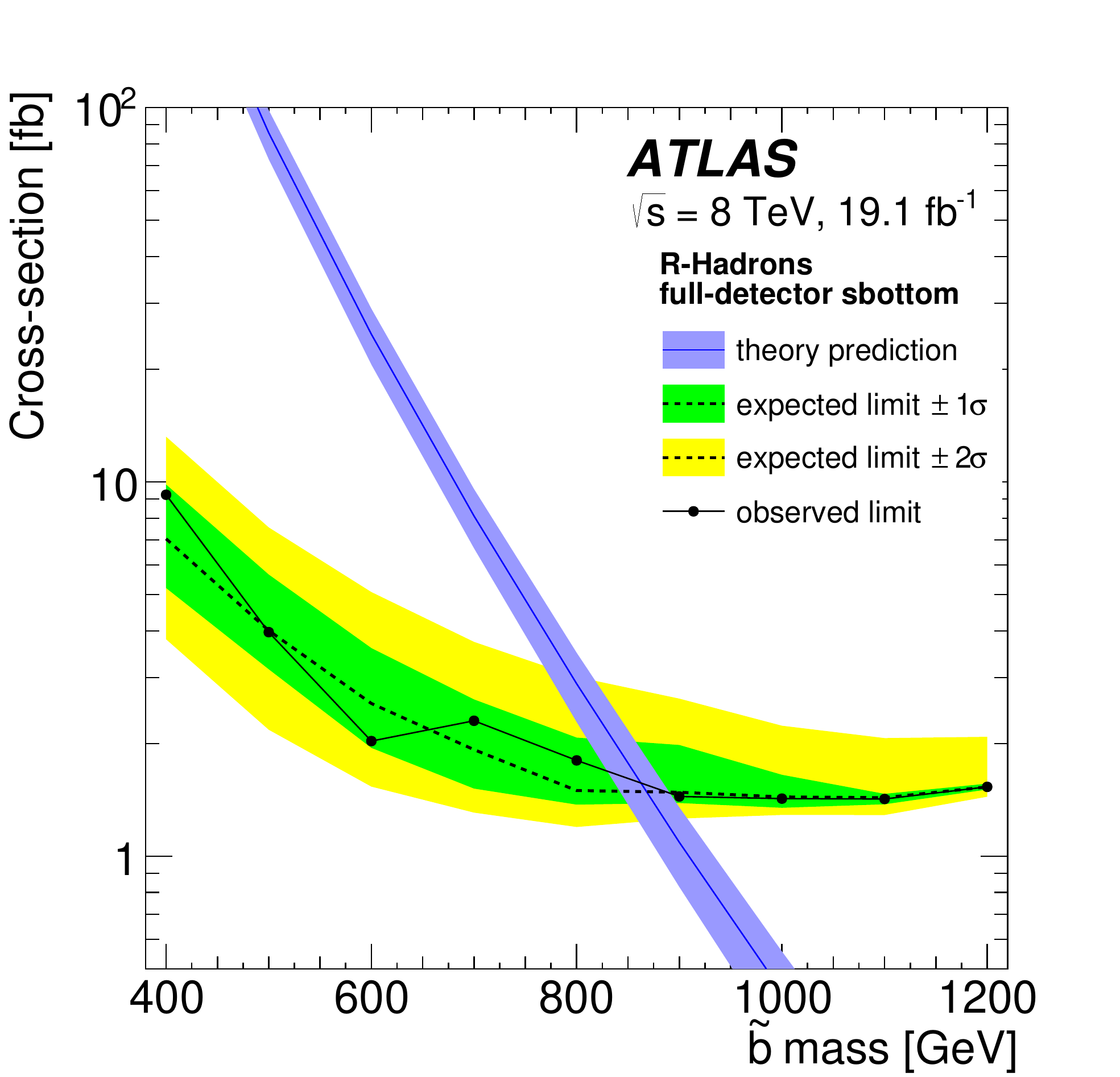}
  \includegraphics[width=0.48\linewidth]{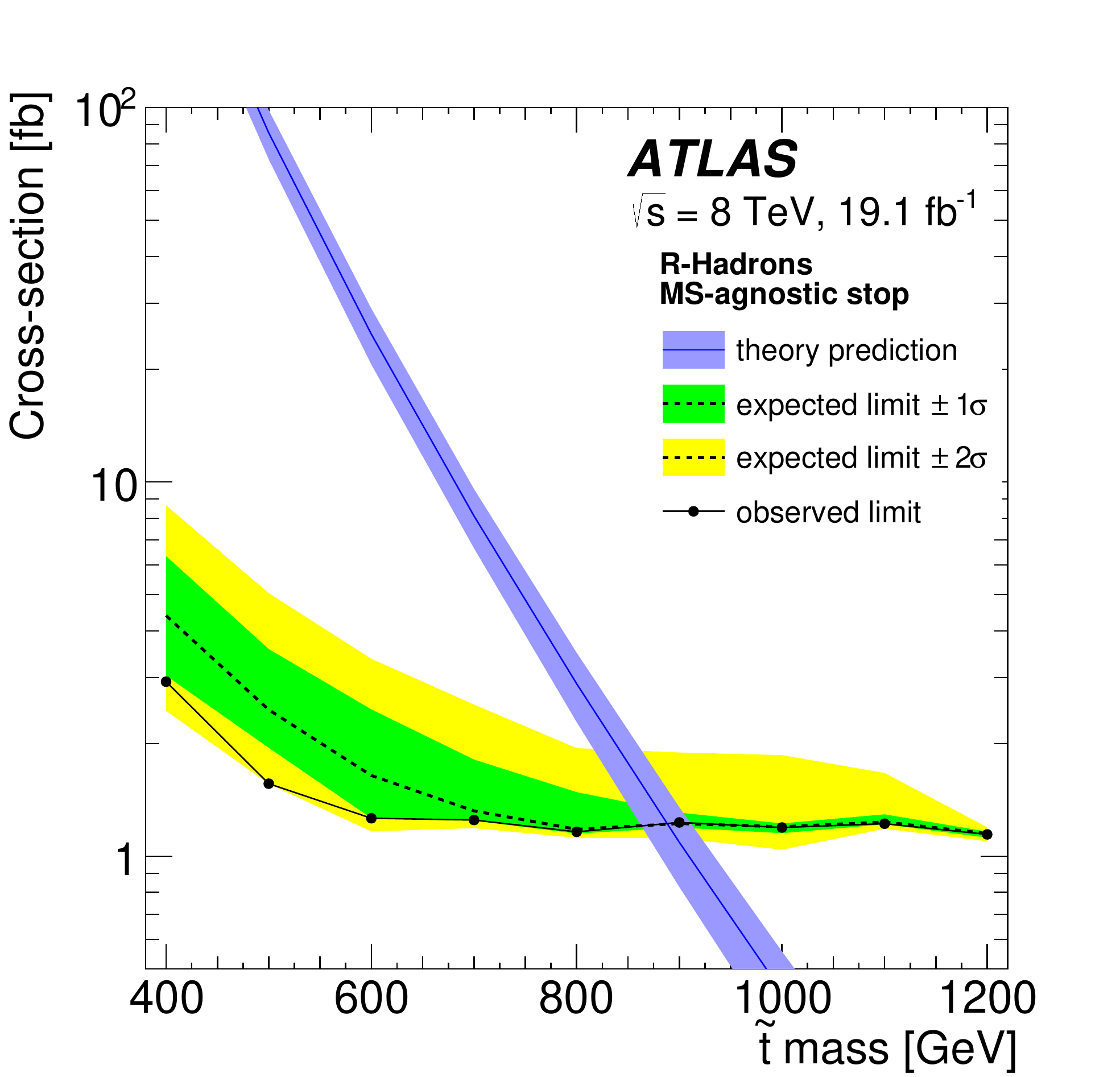}   
  \includegraphics[width=0.48\linewidth]{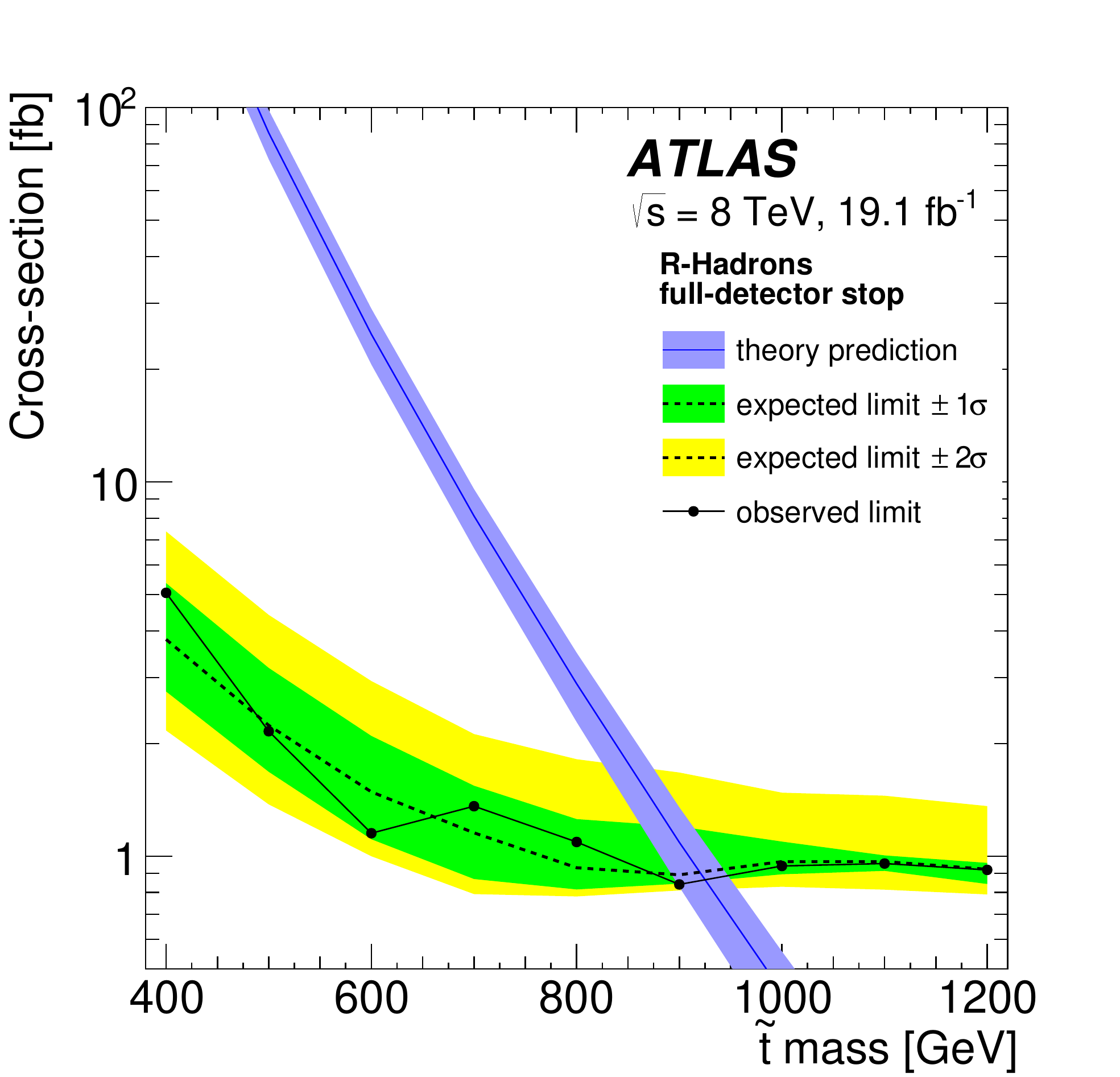}
\caption{Cross-section upper limits as a function of the LLP mass for the \rhad models for the MS-agnostic (left) and full-detector search (right). The expected limit is drawn as a dashed black line with $\pm 1\sigma$ and $\pm 2\sigma$ uncertainty bands drawn in green and yellow, respectively. The observed limit is shown as solid black line with markers. The theoretical cross-section prediction is shown as a solid blue line with a shaded $\pm 1\sigma$ uncertainty band.}
 \label{fig:RHad_limit}
\end{figure*}

The \rhad limits in the MS-agnostic (left) and full-detector (right) searches are shown in figure~\ref{fig:RHad_limit}. In the full-detector search for \rhads a lower mass limit at $95\%$ CL of \limitgluinofull~\GeV for gluinos, \limitsbottomfull~\GeV for sbottoms and \limitstopfull~\GeV for stops is obtained. A selection relying solely on the inner detector and calorimeters, thereby covering e.g.\ \rhads which change into neutral bound states in the calorimeters before reaching the MS, yields a lower mass limit of \limitgluinoagno~\GeV for gluinos, \limitsbottomagno~\GeV for sbottoms and \limitstopagno~\GeV for stops.  

\begin{table}[tb]
  \footnotesize
  \begin{center}
    {\setlength{\tabcolsep}{0em}\begin{tabular*}{\linewidth}{@{\extracolsep{\fill}}lr}
      \hline
      \hline
      Search                                                                             & Lower mass limit [GeV]                                 \\
      \hline
      GMSB sleptons                                                                      &                                                        \\
      ~$\cdot$~$\tan\beta=10,20,30,40,50$                                                & \gmsbAllExclusion                                      \\
      \multicolumn{2}{l}{~$\cdot$~direct $\tilde{\ell}$ production {\scriptsize ($m_{\tilde{\ell}}-m_{\stau_1}=2.7$--93~\GeV)} \hfill 377--335}   \\
      ~$\cdot$~direct $\stau_1$ production                                               & 289                                                    \\
      ~$\cdot$~$\tilde{\chi}^0_1\tilde{\chi}^{\pm}_1$ decaying to stable $\stau_1$       & 537                                                    \\
      \hline
      \lsusy                                                                             &                                                        \\
      ~$\cdot$~$\tilde{q}$,~$\tilde{g}$                                                  & \limitleptosquark, \limitleptogluino                   \\
      \hline
      Charginos                                                                          &                                                        \\
      ~$\cdot$~$\tilde{\chi}^{\pm}_1$                                                    & \limitcharginos                                        \\
      \hline
      \rhads                                                                             &                                                        \\
      ~$\cdot$~$\tilde{g}$, $\tilde{b}$, $\tilde{t}$ (full-detector)                     & \limitgluinofull, \limitsbottomfull and \limitstopfull \\
      ~$\cdot$~$\tilde{g}$, $\tilde{b}$, $\tilde{t}$ (MS-agnostic)                       & \limitgluinoagno, \limitsbottomagno and \limitstopagno \\
      \hline
      \hline 
    \end{tabular*}}
    \caption{Summary of the lower mass limits (95\% CL) from the various searches.} \label{tab:AllResults}
  \end{center}
\end{table}

\section{Conclusion}

Searches for heavy long-lived charged particles are performed through measurement of the mass of candidates by means of time-of-flight and specific ionisation loss measurements in ATLAS sub-detectors using a data sample of \lumi from proton--proton collisions at a centre-of-mass energy of $\sqrt{s} = 8~\TeV$ collected by the ATLAS detector at the Large Hadron Collider. The data are found to match the Standard Model background expectation within uncertainties. The exclusion limits placed for various models impose new constraints on non-SM cross-sections.

An overview on all 95\% CL lower mass limits placed in this article is given in table~\ref{tab:AllResults}.

The upper limits placed on cross-sections and lower limits placed on the mass of long-lived particles in various supersymmetric models, thanks to increased luminosity and more advanced data analysis, substantially extend previous ATLAS limits, and are largely complementary to searches for promptly decaying SUSY particles.

\appendix

\acknowledgments

We thank CERN for the very successful operation of the LHC, as well as the
support staff from our institutions without whom ATLAS could not be
operated efficiently.

We acknowledge the support of ANPCyT, Argentina; YerPhI, Armenia; ARC,
Australia; BMWFW and FWF, Austria; ANAS, Azerbaijan; SSTC, Belarus; CNPq and FAPESP,
Brazil; NSERC, NRC and CFI, Canada; CERN; CONICYT, Chile; CAS, MOST and NSFC,
China; COLCIENCIAS, Colombia; MSMT CR, MPO CR and VSC CR, Czech Republic;
DNRF, DNSRC and Lundbeck Foundation, Denmark; EPLANET, ERC and NSRF, European Union;
IN2P3-CNRS, CEA-DSM/IRFU, France; GNSF, Georgia; BMBF, DFG, HGF, MPG and AvH
Foundation, Germany; GSRT and NSRF, Greece; ISF, MINERVA, GIF, I-CORE and Benoziyo Center,
Israel; INFN, Italy; MEXT and JSPS, Japan; CNRST, Morocco; FOM and NWO,
Netherlands; BRF and RCN, Norway; MNiSW and NCN, Poland; GRICES and FCT, Portugal; MNE/IFA, Romania; MES of Russia and ROSATOM, Russian Federation; JINR; MSTD,
Serbia; MSSR, Slovakia; ARRS and MIZ\v{S}, Slovenia; DST/NRF, South Africa;
MINECO, Spain; SRC and Wallenberg Foundation, Sweden; SER, SNSF and Cantons of
Bern and Geneva, Switzerland; NSC, Taiwan; TAEK, Turkey; STFC, the Royal
Society and Leverhulme Trust, United Kingdom; DOE and NSF, United States of
America.

The crucial computing support from all WLCG partners is acknowledged
gratefully, in particular from CERN and the ATLAS Tier-1 facilities at
TRIUMF (Canada), NDGF (Denmark, Norway, Sweden), CC-IN2P3 (France),
KIT/GridKA (Germany), INFN-CNAF (Italy), NL-T1 (Netherlands), PIC (Spain),
ASGC (Taiwan), RAL (UK) and BNL (USA) and in the Tier-2 facilities
worldwide.

\clearpage

\bibliographystyle{JHEP}
\bibliography{SUSY-2013-22}

\clearpage

\onecolumn
\clearpage 
\begin{flushleft}
{\Large The ATLAS Collaboration}

\bigskip

G.~Aad$^{\rm 85}$,
B.~Abbott$^{\rm 113}$,
J.~Abdallah$^{\rm 152}$,
S.~Abdel~Khalek$^{\rm 117}$,
O.~Abdinov$^{\rm 11}$,
R.~Aben$^{\rm 107}$,
B.~Abi$^{\rm 114}$,
M.~Abolins$^{\rm 90}$,
O.S.~AbouZeid$^{\rm 159}$,
H.~Abramowicz$^{\rm 154}$
% ...
\bigskip
\\
% ...
$^{11}$ Institute of Physics, Azerbaijan Academy of Sciences, Baku, Azerbaijan\\
% ...
$^{85}$ CPPM, Aix-Marseille Universit{\'e} and CNRS/IN2P3, Marseille, France\\
% ...
$^{90}$ Department of Physics and Astronomy, Michigan State University, East Lansing MI, United States of America\\
% ...
$^{107}$ Nikhef National Institute for Subatomic Physics and University of Amsterdam, Amsterdam, Netherlands\\
% ...
$^{113}$ Homer L. Dodge Department of Physics and Astronomy, University of Oklahoma, Norman OK, United States of America\\
$^{114}$ Department of Physics, Oklahoma State University, Stillwater OK, United States of America\\
% ...
$^{117}$ LAL, Universit{\'e} Paris-Sud and CNRS/IN2P3, Orsay, France\\
% ...
$^{152}$ Institute of Physics, Academia Sinica, Taipei, Taiwan\\
% ...
$^{154}$ Raymond and Beverly Sackler School of Physics and Astronomy, Tel Aviv University, Tel Aviv, Israel\\
% ...
$^{159}$ Department of Physics, University of Toronto, Toronto ON, Canada\\
% ...
\end{flushleft}

\end{document}